\documentclass[aps,prl,reprint,amsmath,superscriptaddress]{revtex4-1}
\usepackage{graphicx}
\usepackage{amssymb}
\usepackage[colorlinks,allcolors=blue]{hyperref}
\usepackage[figure,figure*]{hypcap}
\usepackage{soul}

\newcommand{\pref}[2]{\hyperref[#1]{\ref{#1}(#2)}}
\newcommand{\eqpref}[1]{\hyperref[#1]{(\ref{#1})}}

\renewcommand{\section}[1]{{\pdfbookmark[1]{#1}{#1}\vskip\baselineskip\fontfamily{cmss}\fontseries{bx}\selectfont\noindent #1}}
\renewcommand{\subsection}[1]{\pdfbookmark[2]{#1}{#1}\vskip\baselineskip\emph{#1}}

\newcommand{\C}{\mathcal{C}}
\newcommand{\s}{\sigma}

\newcommand{\affUIUC}{\affiliation{Department of Physics, University of Illinois at Urbana-Champaign, Urbana, IL 61801-3080, USA}}
\newcommand{\affICFO}{\affiliation{ICFO--Institut de Ciencies Fotoniques, The Barcelona Institute of Science and Technology, 08860 Castelldefels (Barcelona), Spain}}
\newcommand{\affPolytech}{\affiliation{Departament de F\'isica, Universitat Polit\`ecnica de Catalunya, Campus Nord B4-B5, 08034 Barcelona, Spain}}
\newcommand{\affMariatwo}{\affiliation{Dipartimento di Fisica, Universit\`a di Napoli Federico II, Complesso Universitario di Monte Sant{'}Angelo, Via Cintia, 80126 Napoli, Italy}}

\begin{document}
\title{Observation of the topological Anderson insulator in disordered atomic wires}

\author{Eric J. Meier}
\affUIUC

\author{Fangzhao Alex An}
\affUIUC

\author{Alexandre Dauphin}
\affICFO

\author{Maria Maffei}
\affICFO
\affMariatwo

\author{Pietro Massignan}
\email{pietro.massignan@upc.edu}
\affICFO
\affPolytech

\author{Taylor L. Hughes}
\email{hughest@illinois.edu}
\affUIUC

\author{Bryce Gadway}
\email{bgadway@illinois.edu}
\affUIUC

\date{\today}

\begin{abstract}
Topology and disorder have deep connections and a rich combined influence on quantum transport. In order to probe these connections, we synthesized one-dimensional chiral symmetric wires with controllable disorder via spectroscopic Hamiltonian engineering, based on the laser-driven coupling of discrete momentum states of ultracold atoms. We characterize the system's topology through measurement of the mean chiral displacement of the bulk density extracted from quench dynamics. We find evidence for the topological Anderson insulator phase, in which the band structure of an otherwise trivial wire is driven topological by the presence of added disorder. In addition, we observed the robustness of topological wires to weak disorder and measured the transition to a trivial phase in the presence of strong disorder. Atomic interactions in this quantum simulation platform will enable future realizations of strongly interacting topological fluids.
\end{abstract}

\maketitle

Topology and disorder share many surprising connections, from the formal similarity of one-dimensional (1D) pseudo-disordered lattices and two-dimensional (2D) integer quantum Hall Hofstadter lattices~\cite{Zilb-TopAdiab,Hofst-1976}, to the deep connection between the symmetry classes of random matrices~\cite{Altland-Zirnbauer} and the classification of symmetry protected topological phases~\cite{Ryu-10foldWay}. Recently, there has been great interest in exploring both disorder~\cite{Sanchez-Palencia-Lewenstein-NP-2010} and topology~\cite{Goldman-NatPhysReview} through quantum simulation, stemming from the dramatic influences that these ingredients can have, separately, on the localization properties of quantum particles~\cite{Anderson-1958,Klitzing-IQHE}. When combined, disorder and topology can have a rich and varied influence on quantum transport~\cite{Evers-Mirlin-2008RMP}. Indeed, one of the hallmark features of topological insulators (TIs) is the topologically protected boundary states that are immune to weak disorder~\cite{TI-review}. The robust conductance of such boundary states, such as the 1D edge states of integer quantum Hall systems~\cite{Klitzing-IQHE}, or the 2D surface states of three-dimensional (3D) TIs~\cite{Chen178}, serves as an important counterexample to the inevitability of localization in low-dimensional disordered systems~\cite{Anderson-1958,Abrahams-Scaling-1979}. Despite the robustness to weak disorder, a change in topology can result from strong disorder, and unusual critical phenomena related to the unwinding of the topology can accompany such transitions~\cite{Multifractal-Aoki-1983,Fisher-RandomSinglet}.

Surprisingly, static disorder can also induce nontrivial topology when added to a trivial band structure. This disorder-driven topological phase, known as the \emph{topological Anderson insulator} (TAI), was first predicted to occur in metallic 2D HgTe/CdTe quantum wells~\cite{Li-2009-TopAndIns}. There has been much interest in the TAI phase over the past decade~\cite{Li-2009-TopAndIns,Jiang-2009-NumTAI,Groth-2009-TheoryOfTAI}, and many theoretical studies have shown the TAI phenomenon to be quite general, emerging across a range of disordered systems~\cite{TAIin3D,AltlandCrit,Titum-2015-DisFloquetTI,Liu-Disorder-Top-Photonic}. However, due to the lack of precise control over disorder in real materials, and the difficulty in engineering both topology and disorder in most quantum simulators, the TAI has so far evaded experimental realization.

We engineer synthetic 1D chiral symmetric wires with precisely controllable disorder using simultaneous, coherent control over many transitions between discrete quantum states of ultracold atoms. We directly measure the topological index of the synthetic wires through observation of the bulk dynamics of the atomic density following a quench. We observe a robustness of topological wires to weak tunneling (off-diagonal) disorder, while for very strong disorder we observe a transition from topological to trivial. We furthermore observe that a nontrivial topological band structure, the long-sought TAI phase~\cite{Li-2009-TopAndIns}, can be induced from an incipient non-topological phase through the addition of static disorder. These transitions, enabled by our unique ability to synthesize many precisely controlled disorder realizations, constitute the first detailed investigations of disorder-driven changes in topology in any experimental system.

\begin{figure}
\includegraphics[width=210pt]{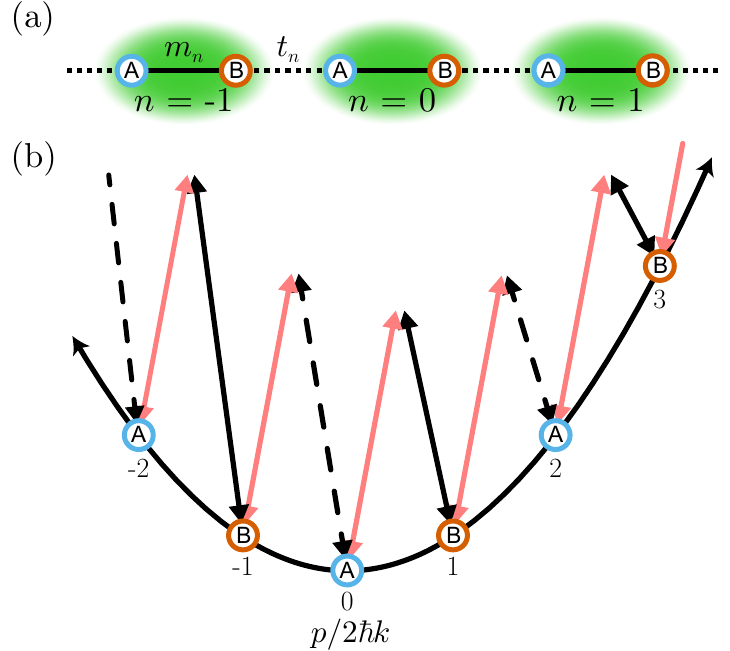}
\caption{Synthetic chiral symmetric wires engineered with atomic momentum states.
\textbf{(a)} Schematic lattice of the nearest-neighbor-coupled chiral symmetric wire. Site-to-site links within the unit cell (solid) and those connecting different unit cells (dashed) have independent tunneling energies $m_n$ and $t_n$, respectively.
\textbf{(b)} Schematic of the experimental implementation of the tight-binding model depicted in \textbf{(a)}, with tunneling based on two-photon Bragg transitions between discrete atomic momentum states.
}
\label{fig1}
\end{figure}

The topological band structures we consider are 1D TIs based off the Su-Schrieffer-Heeger model having chiral, or sublattice, symmetry~\cite{SSH-1979,Ryu-10foldWay,AltlandCrit,DragonAIII}. The rich variety of phenomena associated with 1D TIs (e.g. boundary modes that are stable against disorder, bulk-boundary correspondence, and quantized charge pumping) are easy to visualize in such chiral symmetric wires. We describe this system in terms of a tight-binding model with a two-site unit cell, having sublattice sites $A$ and $B$ [depicted in Fig.~\pref{fig1}{a}]. We consider the Hamiltonian
\begin{equation}
H= \sum_{n} \left[ m_{n} c_n^\dagger S c_n  +  t_{n} \left(c_{n+1}^{\dagger}\frac{(\s_1 - i\s_2)}{2}c_{n} + \textrm{h.c.}\right) \right],
\label{Ham}
\end{equation}
where $c_n^\dagger = ( c_{n,A}^\dagger, c_{n,B}^\dagger )$ creates a particle at unit cell $n$ in sublattice site $A$ or $B,$ $c_n$ is the corresponding annihilation operator, and $\s_i$ are the Pauli matrices related to the sublattice degree of freedom~\cite{DragonAIII}.
The $m_n$ and $t_n$ characterize the intra- and inter-cell tunneling energies. This model can describe chiral wires of the AIII or BDI symmetry classes, by choosing the intra-cell hopping term to be $S=\s_1$ (BDI) or $S=\s_2$ (AIII). Both the AIII and the BDI class models respect chiral symmetry, i.e., they obey $\Gamma H \Gamma = -H$ with $\Gamma = \s_3 \otimes \mathbb{I}$ the chiral operator, whereas the BDI class also obeys particle-hole and time-reversal symmetry~\cite{Ryu-10foldWay}.

We experimentally implement effective tight-binding models of the form of Eq.~\eqref{Ham} using the controlled, parametric coupling of many discrete momentum states of ultracold atoms~\cite{MSL-expt}. We start with a weakly-trapped Bose-Einstein condensate (BEC) of $^{87}$Rb atoms and apply a pair of counter-propagating laser fields with nominal wavelength $\lambda$ and wavevector $k=2\pi/\lambda$. These lasers are far-detuned from any atomic transitions, however their interference pattern couples to the atoms through the ac Stark effect. The spatial periodicity of the laser interference pattern, $\pi/k$, defines the set of momentum states having momenta separated by integer values of $2\hbar k$. These states may be coupled from the BEC, which is a source of atoms with essentially zero momentum,  and they represent the effective sites of our synthetic lattice. The tunneling of atoms between these sites is precisely controlled by simultaneously driving many two-photon Bragg transitions with the applied laser fields. The individual, spectroscopically-resolved control over many such transitions is allowed for by the Doppler shifts experienced by the atoms, which are unique to the various Bragg transitions [depicted in Fig.~\pref{fig1}{b}]. This provides local (in momentum space) control of the intra- and inter-cell tunneling amplitudes and phases, directly through the amplitudes and phases of the corresponding Bragg laser fields~\cite{MSL-expt}.

The simplicity of chiral symmetric wires, described by just a two-site unit cell with separate intra- and inter-cell hoppings, has allowed for several other realizations based on real-space superlattices~\cite{Lohse-pumping,Nakajima-pumping}. Such studies were restricted to exploring BDI wires, as quantum tunneling between stationary lattice sites is real-valued. In comparison, our use of laser-driven tunneling, which allows an independent and arbitrary control of all tunneling phases, gives us access to not only the BDI class but also the AIII class, whose recently proposed~\cite{Velasco-2017-AIII} real-space realization would involve tremendous efforts. Enabled by these controls, we observe a disorder-driven topological to trivial transition in BDI-class wires and also a disorder-driven trivial to topological transition in AIII-class wires, where we find evidence for the TAI phase.

Our ability to create precisely defined disorder in the off-diagonal tunneling terms is crucial for this study. Unlike the site-potential disorder that is more naturally realized in real-space cold atom experiments, e.g., through optical speckle~\cite{Billy-AndersonLocalization-2008} or quasiperiodic lattice potentials~\cite{Roati-AndersonLocalization-2008}, pure tunneling disorder is important for preserving the chiral symmetry of our wires~\cite{AltlandCrit,DragonAIII}. In particular, we let
\begin{align}
t_n &= t(1 + W_1 \omega_n),\\
m_n &= t(m + W_2 \omega_n'),
\end{align}
define the variations of our hopping terms, where $t$ is the characteristic inter-cell tunneling energy, $m$ is the ratio of intra- to inter-cell tunneling in the clean limit, $\omega_n$ and $\omega_n'$ are independent random real numbers chosen uniformly from the range $[-0.5,0.5]$, and $W_1$ and $W_2$ are the dimensionless disorder strengths applied to inter- and intra-cell tunneling.

\begin{figure*}
\includegraphics[width=500pt]{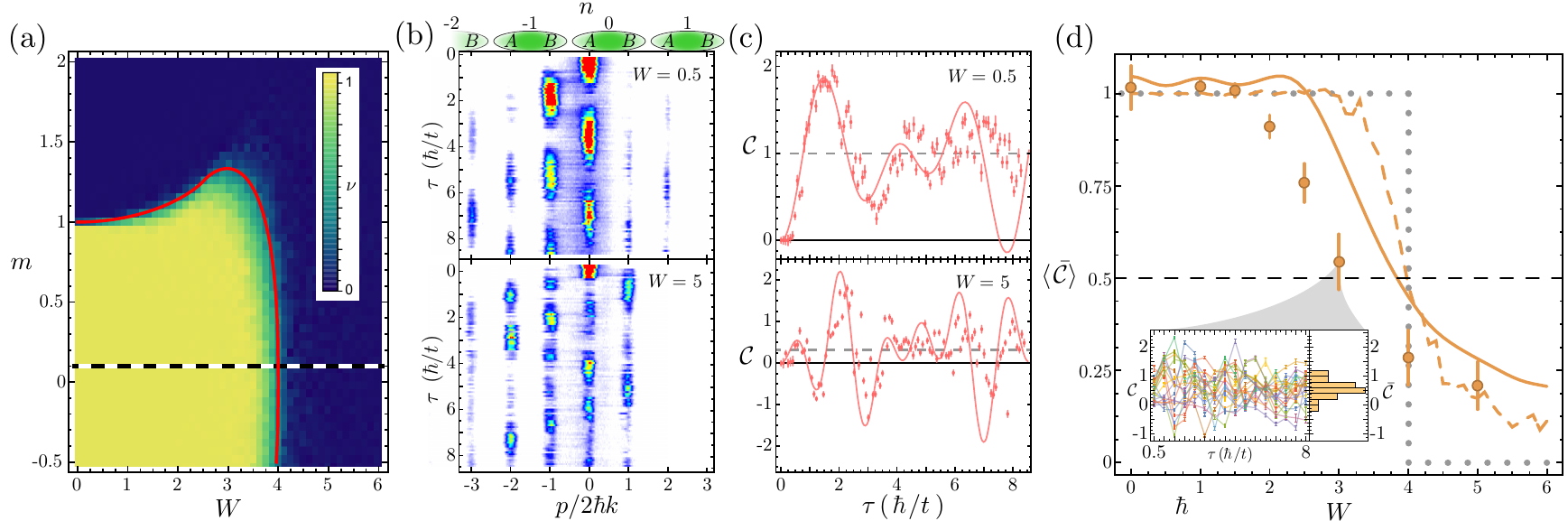}
\caption{Disorder-driven transition from topological to trivial wires.
\textbf{(a)} Topological phase diagram of the BDI wire model described in Eq.~\eqref{Ham}, showing the winding number $\nu$ (inset color scale) as a function of disorder strength $W$ and tunneling ratio $m$ with tunneling disorder strengths $W \equiv W_2 = 2 W_1$. The dashed line at $m=0.1$ indicates the region explored experimentally. The solid red curve indicates the critical phase boundary.
\textbf{(b)} Integrated absorption images of the bulk dynamics following a sudden quench of the tunnel couplings, for both weak disorder ($W = 0.5$) and strong disorder ($W = 5$), each for a single disorder configuration.
\textbf{(c)} Dynamics of $\C$ as calculated from the data shown in \textbf{(b)}. The solid red curves are numerical simulations with no free parameters. The dashed gray horizontal lines denote $\bar{\C}$ for each data set.
\textbf{(d)} $\langle \bar{\C} \rangle$ as a function of $W$ for $m=0.100(5)$. The data are averaged over 20 independent disorder configurations and times in the range 0.5 to $8~\hbar/t$ in steps of $0.5~\hbar/t$. The solid gold line represents a numerical simulation for 200 disorder configurations, but with the same finite time sampling as the data. The dashed gold line is based on the same simulation as the solid gold line, but sampled to much longer times ($\tau = $ 1,000$~\hbar/t$) in a wire with 250 unit cells. The dotted grey curve shows the topological index in the thermodynamic limit~\cite{DragonAIII}, which takes a value of 0.5 at the critical point, as indicated by the horizontal dashed line.
The inset shows $\C$ for $W=3$ as a function of time for all 20 disorder configurations with $\bar{\C}$ for each disorder shown in the histogram.
All error bars in \textbf{(c)} and \textbf{(d)} denote one standard error of the mean.
}
\label{fig2}
\end{figure*}

We begin by considering the influence of disorder added to a BDI-class wire. The wire is strongly dimerized, as characterized by a small intra- to inter-cell tunneling ratio of $m=0.100(5)$ (with $t/\hbar \approx 2\pi \times 1.2$~kHz), and hence is in the topological regime in the clean limit. We fix the disorder amplitudes to be $W \equiv W_2 = 2W_1,$ and show in Fig.~\pref{fig2}{a} the disorder-averaged topological phase diagram of this
model as a function of $W$ and $m$, as determined numerically by a real-space calculation of the winding number $\nu$ for a system with 200 unit cells, together with the critical phase boundary predicted for an infinite system based on the divergence of the localization length~\cite{DragonAIII}.

The strong dimerization produces a large (in units of the bandwidth) energy gap in the band structure. Such large band gaps are typically favorable for experimentally observing the topological nature of disorder-free nontrivial wires via adiabatic charge pumping~\cite{Lohse-pumping,Nakajima-pumping} or the adiabatic preparation of boundary states~\cite{MSL-edge}. However, it is expected that in disordered chiral symmetric wires the bulk energy gap will essentially vanish at moderate disorder strengths, well below those required to induce a change in topology~\cite{DragonAIII}. The energy gap is replaced by a mobility gap, and the band insulator of the clean system is replaced by an Anderson insulator that remains topological, with topology carried by localized states in the spectrum~\cite{DragonAIII}. Thus, without the spectral gap, experimental probes relying on adiabaticity are expected to fail in evidencing the topology of disordered wires.

We instead characterize the topology of our wires by monitoring the bulk dynamical response of atoms to a sudden quench. Specifically, we measure the \emph{mean chiral displacement} (MCD) of our atoms. This observable was recently introduced in the context of discrete-time photonic quantum walks~\cite{photonicCD}, and is herein measured for the first time for continuous-time dynamics. We define the expectation value of the chiral displacement operator as
\begin{equation}
\C  = 2 \langle\Gamma X\rangle \ ,
\end{equation}
given in terms of the chiral operator $\Gamma$ and the unit cell operator $X$~\cite{photonicCD}. The dynamics of $\C$ in general display a transient, oscillatory behavior, and its time- and disorder-average $\langle \bar{\C} \rangle$ converges to the winding number $\nu$, or equivalently to the Zak phase $\varphi_\mathrm{Zak}$ divided by $\pi$, in both the clean and the disordered cases.
At topological critical points, moreover, $\langle \bar{\C} \rangle$ converges to the average of the invariants computed in the two neighboring phases~\cite{photonicCD,MariaNJP,Supplement}.

For our experiment we begin with all tunnel couplings turned off, and the entire atomic population localized at a single central bulk lattice site (site $A$ of unit cell $n=0$, for a system with 20 unit cells). We then quench on the tunnel couplings in a stepwise fashion. The projection of the localized initial state onto the quenched system's eigenstates leads to rich dynamics, as depicted in Fig.~\pref{fig2}{b} for both weak ($W=0.5$) and strong ($W = 5$) disorder. Such site-resolved dynamics of the atomic population distribution are directly measured by a series of absorption images taken after dynamical evolution under the Hamiltonian of Eq.~\eqref{Ham} for a variable time $\tau$ (given in units of the tunneling time $\hbar/t \approx 130$~$\mu$s), and after the discrete momentum states separate according to their momenta during a time-of-flight period~\cite{MSL-expt}. From the data shown in Fig.~\pref{fig2}{b} we calculate $\C$ as a function of $\tau$ as shown in Fig.~\pref{fig2}{c}, along with the time average ${\bar{\C}}$. We additionally consider the disorder-averaged topological characterization of the system by averaging $\bar{\C}$ over many independent disorder configurations.

\begin{figure*}
\includegraphics[width=436pt]{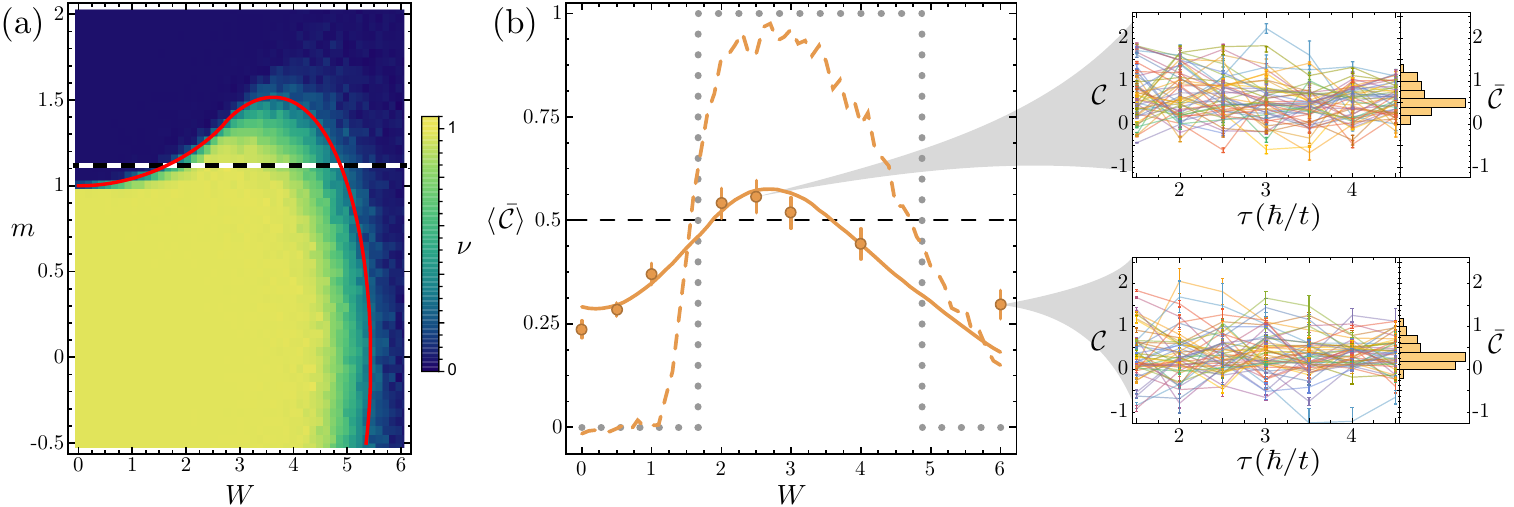}
\caption{Observation of the topological Anderson insulator phase.
\textbf{(a)} Topological phase diagram of the AIII wire model described in Eq.~\eqref{Ham}, showing the computed winding number (color scale at right) as a function of disorder strength $W$ and tunneling ratio $m$ with tunneling disorder strengths $W \equiv W_2$ ($W_1 = 0$). The striped black and white line at $m = 1.12$ indicates the region explored experimentally. The solid red curve indicates the critical boundary (i.e. the set of points where the localization length diverges for an infinite chain).
\textbf{(b)} $\langle \bar{\C} \rangle$ as a function of $W$ for $m=1.12(2)$. The data are averaged over 50 independent disorder configurations and are averaged in time over the range 1.5 to $4.5~\hbar/t$ in steps of $0.5~\hbar/t$. The solid gold line relates to a numerical simulation for 200 disorder configurations, but with the same finite time sampling as the data. The dashed gold line is based on the same simulation as the solid gold line, but sampled to much longer times ($\tau =$ 1,000 $\hbar/t$) in a 250 unit cell system.
The dotted grey curve shows the topological index in the thermodynamic limit~\cite{DragonAIII}, which takes a value of 0.5 at the critical points, as indicated by the horizontal dashed line.
$\C$ as a function of time for all 50 disorder realizations is shown at right for $W=2.5$ and $6$ and $\bar{\C}$ are shown in the histogram to the right of each plot. All error bars in \textbf{(b)} denote one standard error of the mean.
}
\label{fig3}
\end{figure*}

The dependence of $\langle\bar{\C}\rangle$ on the strength of applied disorder $W$ is summarized in Fig.~\pref{fig2}{d}. The inset of Fig.~\pref{fig2}{d} depicts the determination of $\langle \bar{\C} \rangle$ (shown for the case $W=3$), first from the time-average of $\C$ over 16 values of $\tau$ evenly spaced between $0.5~\hbar/t$ and $8~\hbar/t$, followed by an average over 20 unique realizations of disorder. We observe a robustness of $\langle \bar{\C} \rangle$ to weak disorder, maintaining a nearly-quantized value close to one. For strong disorder, $W \gtrsim 2$, we observe a relatively steep drop in $\langle \bar{\C} \rangle$, with it falling below $\langle \bar{\C} \rangle=0.5$ for $W \gtrsim 3$. Our observed decrease of $\langle \bar{\C} \rangle$ with increasing disorder is in good agreement with a numerical simulation (solid gold line) of the Hamiltonian in Eq.~\eqref{Ham} for experimental time scales. The observed decay of $\langle \bar{\C} \rangle$ is associated to a disorder-driven transition between topological ($W \lesssim 4$) and trivial wires ($W \gtrsim 4$). This observed crossover represents the first systematic exploration of a disorder-driven change of topology in any system, enabled by our unique ability to precisely define disorder configurations.

On an infinitely long chain, we would expect to observe a sharp phase transition in the infinite-time limit of our $\langle \bar{\C} \rangle$ measurement, yielding quantized values of the invariant for all disorders, and half-integer values at the critical phase boundary~\cite{Altland-PRB2015,Supplement}. However, we instead observe a smooth crossover due to finite-time broadening from our abbreviated period of quench dynamics and the corresponding finite number of sites. The observation of a moderately sharper transition, such as that of the dashed-line numerical simulation in Fig.~\pref{fig2}{d}, would require that we measure at extremely long timescales (shown for 1,000 tunneling times) and for very large systems (shown for 250 unit cells), which at the moment is beyond the capabilities of our experimental technique. The slow convergence of this transition with increasing measurement time and system size is a characteristic feature of random-singlet transitions~\cite{Fisher-RandomSinglet}, such as that found in chiral symmetric wires at strong disorder.

Having demonstrated a disorder-driven change of topology in BDI-class wires, we now turn our attention to AIII-class wires where we investigate the surprising feature that an initially clean, trivial system can be driven topological through the addition of disorder. This phenomenon is manifest in the phase diagram of AIII-class wires in Fig.~\pref{fig3}{a} for $m$ just exceeding one. The value $|m|=1$ is the critical point between the topological and trivial phase in the clean limit, and values of $|m|>1$ are in the trivial phase in the absence of disorder. However, we see that random tunneling disorder induces the TAI phase over a broad range of weak to moderate $W$ values, eventually giving way to a trivial Anderson insulator phase again for very large disorder. Beyond numerics, a mechanism for the formation of a TAI phase was first elaborated in Ref.~\cite{Groth-2009-TheoryOfTAI} for 2D systems. In that work, disorder is taken into account perturbatively using the self-consistent Born approximation, and was shown to effectively renormalize the parameters in the Hamiltonian (including the parameter(s) that tune between the topological and trivial phases). The TAI phase arises because, as disorder is added to the trivial phase tuned near the clean critical point, the effective Hamiltonian is renormalized through the critical point and into the topological phase. This type of reasoning was adapted and extended to describe the TAI phase in 1D systems including both the BDI- and AIII-class wires that we consider here~\cite{AltlandCrit,Altland-PRB2015}. Although, strictly speaking, the sharpest results of the latter work can only be applied in a scenario with multiple wires, numerical evidence supports the existence of the TAI even for a single wire~\cite{DragonAIII}.

Here, we probe the influence of tunneling disorder on atomic wires of the AIII class. Since we are interested in the TAI phase, we start with a slight dimerization [$m = 1.12(2)$] that places the system in a trivial phase in the clean limit. We note that being so near the critical point at $m=1$ causes the band gap in the clean limit to be much smaller than in the previous experimental setup. The choice of disorder we consider here differs from the previous case: we add disorder only to the intra-cell hopping terms, i.e., setting $W_1 = 0$ and $W\equiv W_2$. From Refs.~\cite{Groth-2009-TheoryOfTAI,AltlandCrit,Altland-PRB2015} we expect that, for weak disorder of this form, the intra-cell hopping $m$ should be renormalized toward the topological phase resulting in a TAI. Due to the smaller band gap in this case of reduced dimerization, the effects of residual time dependence in our driven system become more important. In order to mitigate these effects, we reduce our tunneling energy to $t/\hbar \approx 2\pi \times 600$~Hz, resulting in a correspondingly lessened experimental time range of $\tau = 1.5$~$\hbar/t$ to 4.5~$\hbar/t$.

Figure~\pref{fig3}{b} shows the dependence of $\langle\bar{\C}\rangle$ on the strength of added disorder in the AIII-class wire. The measured $\langle\bar{\C}\rangle$ values are obtained, as in the previous case, through the non-equilibrium bulk dynamics of the atoms following a quench of the tunneling. Due to the restricted range of $\tau$, we include many more disorder configurations (50) to allow for stable measures of $\langle\bar{\C}\rangle$. For weak disorder, we observe that $\langle\bar{\C}\rangle$ rises and reaches a pronounced maximum at $W \approx 2.5$. This is consistent with the expected change in the renormalized $m$ parameter, i.e., given the negative sign of the lowest-order correction to $m$, for weak disorder \cite{Groth-2009-TheoryOfTAI,AltlandCrit,Altland-PRB2015}. $\langle\bar{\C}\rangle$ then decays for very strong applied disorder. This observation of an initial increase of $\langle\bar{\C}\rangle$ followed by a decrease is indicative of two phase transitions, first from trivial wires to the TAI phase and then to a trivial Anderson insulator at strong disorder, broadened by our finite interrogation time.

Despite the effects of finite-time broadening, we see our measured $\langle\bar{\C}\rangle$ rise to greater than 0.5 (the infinite-time $\langle\bar{\C}\rangle$ value associated with the critical point) for $W\approx 2.5$, lending further evidence to our observation of the TAI phase. The excellent agreement of our experimental $\langle\bar{\C}\rangle$ data with a short-time sampled numerical simulation (solid gold line), combined with the sharper transitions expected for long-time measurements based on the same simulations (dashed gold line for 1,000 tunneling times in a 250 unit cell system), provide strong evidence for the observation of disorder-driven topology in an otherwise trivial band structure.

Unlike with real condensed matter systems and photonic simulators, where carrier mobility or lattice parameters vary from sample to sample, the spectroscopic control of our atomic physics platform has allowed us to engineer many different, precisely tuned realizations of disorder. In addition to enabling these first explorations of disorder-driven quantum phase transitions in topological wires, this unique level of control will also enable future studies of quantum criticality in disordered topological systems~\cite{Evers-Mirlin-2008RMP,Multifractal-Aoki-1983}. By simple extension to longer evolution times, we may study in detail the interesting physics of logarithmic delocalization at the random singlet transition~\cite{Fisher-RandomSinglet}. Combined with our ability to engineer tunneling phases~\cite{MSL-expt} and artificial gauge fields, our technique may be extended to study disordered quantum Hall systems~\cite{Multifractal-Aoki-1983}. And while our present study has been restricted to a regime where interactions are relatively unimportant, the presence of strong interactions in synthetic momentum-space lattices~\cite{MSL-interactions} will enable future studies of strongly interacting topological fluids.

\section{Acknowledgments}
We thank Nathan Goldman, Maciej Lewenstein, Hassan Shapourian, and Ian Mondragon--Shem for helpful discussions. This material is based upon work supported by the National Science Foundation under Grant No. PHY17-07731 (EJM, FAA, and BG). AD, MM, and PM acknowledge Spanish MINECO (Severo Ochoa SEV-2015-0522, FisicaTeAMO FIS2016-79508-P, and SWUQM FIS2017-84114-C2-1-P), the Generalitat de Catalunya (SGR874 and CERCA), the EU (ERC AdG OSYRIS 339106, H2020-FETProAct QUIC 641122), the Fundaci\'o Privada Cellex, a Cellex-ICFO-MPQ fellowship, and the ``Ram\'on y Cajal'' program. TLH was supported by the ONR YIP Award N00014-15-1-2383.


\begin{thebibliography}{35}%
\makeatletter
\providecommand \@ifxundefined [1]{%
 \@ifx{#1\undefined}
}%
\providecommand \@ifnum [1]{%
 \ifnum #1\expandafter \@firstoftwo
 \else \expandafter \@secondoftwo
 \fi
}%
\providecommand \@ifx [1]{%
 \ifx #1\expandafter \@firstoftwo
 \else \expandafter \@secondoftwo
 \fi
}%
\providecommand \natexlab [1]{#1}%
\providecommand \enquote  [1]{``#1''}%
\providecommand \bibnamefont  [1]{#1}%
\providecommand \bibfnamefont [1]{#1}%
\providecommand \citenamefont [1]{#1}%
\providecommand \href@noop [0]{\@secondoftwo}%
\providecommand \href [0]{\begingroup \@sanitize@url \@href}%
\providecommand \@href[1]{\@@startlink{#1}\@@href}%
\providecommand \@@href[1]{\endgroup#1\@@endlink}%
\providecommand \@sanitize@url [0]{\catcode `\\12\catcode `\$12\catcode
  `\&12\catcode `\#12\catcode `\^12\catcode `\_12\catcode `\%12\relax}%
\providecommand \@@startlink[1]{}%
\providecommand \@@endlink[0]{}%
\providecommand \url  [0]{\begingroup\@sanitize@url \@url }%
\providecommand \@url [1]{\endgroup\@href {#1}{\urlprefix }}%
\providecommand \urlprefix  [0]{URL }%
\providecommand \Eprint [0]{\href }%
\providecommand \doibase [0]{http://dx.doi.org/}%
\providecommand \selectlanguage [0]{\@gobble}%
\providecommand \bibinfo  [0]{\@secondoftwo}%
\providecommand \bibfield  [0]{\@secondoftwo}%
\providecommand \translation [1]{[#1]}%
\providecommand \BibitemOpen [0]{}%
\providecommand \bibitemStop [0]{}%
\providecommand \bibitemNoStop [0]{.\EOS\space}%
\providecommand \EOS [0]{\spacefactor3000\relax}%
\providecommand \BibitemShut  [1]{\csname bibitem#1\endcsname}%
\let\auto@bib@innerbib\@empty
\bibitem [{\citenamefont {Kraus}\ \emph {et~al.}(2012)\citenamefont {Kraus},
  \citenamefont {Lahini}, \citenamefont {Ringel}, \citenamefont {Verbin},\ and\
  \citenamefont {Zilberberg}}]{Zilb-TopAdiab}%
  \BibitemOpen
  \bibfield  {author} {\bibinfo {author} {\bibfnamefont {Y.~E.}\ \bibnamefont
  {Kraus}}, \bibinfo {author} {\bibfnamefont {Y.}~\bibnamefont {Lahini}},
  \bibinfo {author} {\bibfnamefont {Z.}~\bibnamefont {Ringel}}, \bibinfo
  {author} {\bibfnamefont {M.}~\bibnamefont {Verbin}}, \ and\ \bibinfo {author}
  {\bibfnamefont {O.}~\bibnamefont {Zilberberg}},\ }\href {\doibase
  10.1103/PhysRevLett.109.106402} {\bibfield  {journal} {\bibinfo  {journal}
  {Phys. Rev. Lett.}\ }\textbf {\bibinfo {volume} {109}},\ \bibinfo {pages}
  {106402} (\bibinfo {year} {2012})}\BibitemShut {NoStop}%
\bibitem [{\citenamefont {Hofstadter}(1976)}]{Hofst-1976}%
  \BibitemOpen
  \bibfield  {author} {\bibinfo {author} {\bibfnamefont {D.~R.}\ \bibnamefont
  {Hofstadter}},\ }\href {\doibase 10.1103/PhysRevB.14.2239} {\bibfield
  {journal} {\bibinfo  {journal} {Phys. Rev. B}\ }\textbf {\bibinfo {volume}
  {14}},\ \bibinfo {pages} {2239} (\bibinfo {year} {1976})}\BibitemShut
  {NoStop}%
\bibitem [{\citenamefont {Altland}\ and\ \citenamefont
  {Zirnbauer}(1997)}]{Altland-Zirnbauer}%
  \BibitemOpen
  \bibfield  {author} {\bibinfo {author} {\bibfnamefont {A.}~\bibnamefont
  {Altland}}\ and\ \bibinfo {author} {\bibfnamefont {M.~R.}\ \bibnamefont
  {Zirnbauer}},\ }\href {\doibase 10.1103/PhysRevB.55.1142} {\bibfield
  {journal} {\bibinfo  {journal} {Phys. Rev. B}\ }\textbf {\bibinfo {volume}
  {55}},\ \bibinfo {pages} {1142} (\bibinfo {year} {1997})}\BibitemShut
  {NoStop}%
\bibitem [{\citenamefont {Ryu}\ \emph {et~al.}(2010)\citenamefont {Ryu},
  \citenamefont {Schnyder}, \citenamefont {Furusaki},\ and\ \citenamefont
  {Ludwig}}]{Ryu-10foldWay}%
  \BibitemOpen
  \bibfield  {author} {\bibinfo {author} {\bibfnamefont {S.}~\bibnamefont
  {Ryu}}, \bibinfo {author} {\bibfnamefont {A.~P.}\ \bibnamefont {Schnyder}},
  \bibinfo {author} {\bibfnamefont {A.}~\bibnamefont {Furusaki}}, \ and\
  \bibinfo {author} {\bibfnamefont {A.~W.~W.}\ \bibnamefont {Ludwig}},\ }\href
  {http://stacks.iop.org/1367-2630/12/i=6/a=065010} {\bibfield  {journal}
  {\bibinfo  {journal} {New J. Phys.}\ }\textbf {\bibinfo {volume} {12}},\
  \bibinfo {pages} {065010} (\bibinfo {year} {2010})}\BibitemShut {NoStop}%
\bibitem [{\citenamefont {Sanchez-Palencia}\ and\ \citenamefont
  {Lewenstein}(2010)}]{Sanchez-Palencia-Lewenstein-NP-2010}%
  \BibitemOpen
  \bibfield  {author} {\bibinfo {author} {\bibfnamefont {L.}~\bibnamefont
  {Sanchez-Palencia}}\ and\ \bibinfo {author} {\bibfnamefont {M.}~\bibnamefont
  {Lewenstein}},\ }\href {\doibase 10.1038/nphys1507} {\bibfield  {journal}
  {\bibinfo  {journal} {Nat. Phys.}\ }\textbf {\bibinfo {volume} {6}},\
  \bibinfo {pages} {87} (\bibinfo {year} {2010})}\BibitemShut {NoStop}%
\bibitem [{\citenamefont {Goldman}\ \emph {et~al.}(2016)\citenamefont
  {Goldman}, \citenamefont {Budich},\ and\ \citenamefont
  {Zoller}}]{Goldman-NatPhysReview}%
  \BibitemOpen
  \bibfield  {author} {\bibinfo {author} {\bibfnamefont {N.}~\bibnamefont
  {Goldman}}, \bibinfo {author} {\bibfnamefont {J.~C.}\ \bibnamefont {Budich}},
  \ and\ \bibinfo {author} {\bibfnamefont {P.}~\bibnamefont {Zoller}},\ }\href
  {\doibase 10.1038/nphys3803} {\bibfield  {journal} {\bibinfo  {journal} {Nat.
  Phys.}\ }\textbf {\bibinfo {volume} {12}},\ \bibinfo {pages} {639} (\bibinfo
  {year} {2016})}\BibitemShut {NoStop}%
\bibitem [{\citenamefont {Anderson}(1958)}]{Anderson-1958}%
  \BibitemOpen
  \bibfield  {author} {\bibinfo {author} {\bibfnamefont {P.~W.}\ \bibnamefont
  {Anderson}},\ }\href {\doibase 10.1103/PhysRev.109.1492} {\bibfield
  {journal} {\bibinfo  {journal} {Phys. Rev.}\ }\textbf {\bibinfo {volume}
  {109}},\ \bibinfo {pages} {1492} (\bibinfo {year} {1958})}\BibitemShut
  {NoStop}%
\bibitem [{\citenamefont {von Klitzing}\ \emph {et~al.}(1980)\citenamefont {von
  Klitzing}, \citenamefont {Dorda},\ and\ \citenamefont
  {Pepper}}]{Klitzing-IQHE}%
  \BibitemOpen
  \bibfield  {author} {\bibinfo {author} {\bibfnamefont {K.}~\bibnamefont {von
  Klitzing}}, \bibinfo {author} {\bibfnamefont {G.}~\bibnamefont {Dorda}}, \
  and\ \bibinfo {author} {\bibfnamefont {M.}~\bibnamefont {Pepper}},\ }\href
  {\doibase 10.1103/PhysRevLett.45.494} {\bibfield  {journal} {\bibinfo
  {journal} {Phys. Rev. Lett.}\ }\textbf {\bibinfo {volume} {45}},\ \bibinfo
  {pages} {494} (\bibinfo {year} {1980})}\BibitemShut {NoStop}%
\bibitem [{\citenamefont {Evers}\ and\ \citenamefont
  {Mirlin}(2008)}]{Evers-Mirlin-2008RMP}%
  \BibitemOpen
  \bibfield  {author} {\bibinfo {author} {\bibfnamefont {F.}~\bibnamefont
  {Evers}}\ and\ \bibinfo {author} {\bibfnamefont {A.~D.}\ \bibnamefont
  {Mirlin}},\ }\href {\doibase 10.1103/RevModPhys.80.1355} {\bibfield
  {journal} {\bibinfo  {journal} {Rev. Mod. Phys.}\ }\textbf {\bibinfo {volume}
  {80}},\ \bibinfo {pages} {1355} (\bibinfo {year} {2008})}\BibitemShut
  {NoStop}%
\bibitem [{\citenamefont {Qi}\ and\ \citenamefont {Zhang}(2011)}]{TI-review}%
  \BibitemOpen
  \bibfield  {author} {\bibinfo {author} {\bibfnamefont {X.-L.}\ \bibnamefont
  {Qi}}\ and\ \bibinfo {author} {\bibfnamefont {S.-C.}\ \bibnamefont {Zhang}},\
  }\href {\doibase 10.1103/RevModPhys.83.1057} {\bibfield  {journal} {\bibinfo
  {journal} {Rev. Mod. Phys.}\ }\textbf {\bibinfo {volume} {83}},\ \bibinfo
  {pages} {1057} (\bibinfo {year} {2011})}\BibitemShut {NoStop}%
\bibitem [{\citenamefont {Chen}\ \emph {et~al.}(2009)\citenamefont {Chen},
  \citenamefont {Analytis}, \citenamefont {Chu}, \citenamefont {Liu},
  \citenamefont {Mo}, \citenamefont {Qi}, \citenamefont {Zhang}, \citenamefont
  {Lu}, \citenamefont {Dai}, \citenamefont {Fang}, \citenamefont {Zhang},
  \citenamefont {Fisher}, \citenamefont {Hussain},\ and\ \citenamefont
  {Shen}}]{Chen178}%
  \BibitemOpen
  \bibfield  {author} {\bibinfo {author} {\bibfnamefont {Y.~L.}\ \bibnamefont
  {Chen}}, \bibinfo {author} {\bibfnamefont {J.~G.}\ \bibnamefont {Analytis}},
  \bibinfo {author} {\bibfnamefont {J.-H.}\ \bibnamefont {Chu}}, \bibinfo
  {author} {\bibfnamefont {Z.~K.}\ \bibnamefont {Liu}}, \bibinfo {author}
  {\bibfnamefont {S.-K.}\ \bibnamefont {Mo}}, \bibinfo {author} {\bibfnamefont
  {X.~L.}\ \bibnamefont {Qi}}, \bibinfo {author} {\bibfnamefont {H.~J.}\
  \bibnamefont {Zhang}}, \bibinfo {author} {\bibfnamefont {D.~H.}\ \bibnamefont
  {Lu}}, \bibinfo {author} {\bibfnamefont {X.}~\bibnamefont {Dai}}, \bibinfo
  {author} {\bibfnamefont {Z.}~\bibnamefont {Fang}}, \bibinfo {author}
  {\bibfnamefont {S.~C.}\ \bibnamefont {Zhang}}, \bibinfo {author}
  {\bibfnamefont {I.~R.}\ \bibnamefont {Fisher}}, \bibinfo {author}
  {\bibfnamefont {Z.}~\bibnamefont {Hussain}}, \ and\ \bibinfo {author}
  {\bibfnamefont {Z.-X.}\ \bibnamefont {Shen}},\ }\href {\doibase
  10.1126/science.1173034} {\bibfield  {journal} {\bibinfo  {journal}
  {Science}\ }\textbf {\bibinfo {volume} {325}},\ \bibinfo {pages} {178}
  (\bibinfo {year} {2009})}\BibitemShut {NoStop}%
\bibitem [{\citenamefont {Abrahams}\ \emph {et~al.}(1979)\citenamefont
  {Abrahams}, \citenamefont {Anderson}, \citenamefont {Licciardello},\ and\
  \citenamefont {Ramakrishnan}}]{Abrahams-Scaling-1979}%
  \BibitemOpen
  \bibfield  {author} {\bibinfo {author} {\bibfnamefont {E.}~\bibnamefont
  {Abrahams}}, \bibinfo {author} {\bibfnamefont {P.~W.}\ \bibnamefont
  {Anderson}}, \bibinfo {author} {\bibfnamefont {D.~C.}\ \bibnamefont
  {Licciardello}}, \ and\ \bibinfo {author} {\bibfnamefont {T.~V.}\
  \bibnamefont {Ramakrishnan}},\ }\href {\doibase 10.1103/PhysRevLett.42.673}
  {\bibfield  {journal} {\bibinfo  {journal} {Phys. Rev. Lett.}\ }\textbf
  {\bibinfo {volume} {42}},\ \bibinfo {pages} {673} (\bibinfo {year}
  {1979})}\BibitemShut {NoStop}%
\bibitem [{\citenamefont {Aoki}(1983)}]{Multifractal-Aoki-1983}%
  \BibitemOpen
  \bibfield  {author} {\bibinfo {author} {\bibfnamefont {H.}~\bibnamefont
  {Aoki}},\ }\href {\doibase 10.1088/0022-3719/16/6/007} {\bibfield  {journal}
  {\bibinfo  {journal} {J. Phys. C: Solid State Phys.}\ }\textbf {\bibinfo
  {volume} {16}},\ \bibinfo {pages} {L205} (\bibinfo {year}
  {1983})}\BibitemShut {NoStop}%
\bibitem [{\citenamefont {Fisher}(1994)}]{Fisher-RandomSinglet}%
  \BibitemOpen
  \bibfield  {author} {\bibinfo {author} {\bibfnamefont {D.~S.}\ \bibnamefont
  {Fisher}},\ }\href {\doibase 10.1103/PhysRevB.50.3799} {\bibfield  {journal}
  {\bibinfo  {journal} {Phys. Rev. B}\ }\textbf {\bibinfo {volume} {50}},\
  \bibinfo {pages} {3799} (\bibinfo {year} {1994})}\BibitemShut {NoStop}%
\bibitem [{\citenamefont {Li}\ \emph {et~al.}(2009)\citenamefont {Li},
  \citenamefont {Chu}, \citenamefont {Jain},\ and\ \citenamefont
  {Shen}}]{Li-2009-TopAndIns}%
  \BibitemOpen
  \bibfield  {author} {\bibinfo {author} {\bibfnamefont {J.}~\bibnamefont
  {Li}}, \bibinfo {author} {\bibfnamefont {R.-L.}\ \bibnamefont {Chu}},
  \bibinfo {author} {\bibfnamefont {J.~K.}\ \bibnamefont {Jain}}, \ and\
  \bibinfo {author} {\bibfnamefont {S.-Q.}\ \bibnamefont {Shen}},\ }\href
  {\doibase 10.1103/PhysRevLett.102.136806} {\bibfield  {journal} {\bibinfo
  {journal} {Phys. Rev. Lett.}\ }\textbf {\bibinfo {volume} {102}},\ \bibinfo
  {pages} {136806} (\bibinfo {year} {2009})}\BibitemShut {NoStop}%
\bibitem [{\citenamefont {Jiang}\ \emph {et~al.}(2009)\citenamefont {Jiang},
  \citenamefont {Wang}, \citenamefont {Sun},\ and\ \citenamefont
  {Xie}}]{Jiang-2009-NumTAI}%
  \BibitemOpen
  \bibfield  {author} {\bibinfo {author} {\bibfnamefont {H.}~\bibnamefont
  {Jiang}}, \bibinfo {author} {\bibfnamefont {L.}~\bibnamefont {Wang}},
  \bibinfo {author} {\bibfnamefont {Q.-f.}\ \bibnamefont {Sun}}, \ and\
  \bibinfo {author} {\bibfnamefont {X.~C.}\ \bibnamefont {Xie}},\ }\href
  {\doibase 10.1103/PhysRevB.80.165316} {\bibfield  {journal} {\bibinfo
  {journal} {Phys. Rev. B}\ }\textbf {\bibinfo {volume} {80}},\ \bibinfo
  {pages} {165316} (\bibinfo {year} {2009})}\BibitemShut {NoStop}%
\bibitem [{\citenamefont {Groth}\ \emph {et~al.}(2009)\citenamefont {Groth},
  \citenamefont {Wimmer}, \citenamefont {Akhmerov}, \citenamefont
  {Tworzyd\l{}o},\ and\ \citenamefont {Beenakker}}]{Groth-2009-TheoryOfTAI}%
  \BibitemOpen
  \bibfield  {author} {\bibinfo {author} {\bibfnamefont {C.~W.}\ \bibnamefont
  {Groth}}, \bibinfo {author} {\bibfnamefont {M.}~\bibnamefont {Wimmer}},
  \bibinfo {author} {\bibfnamefont {A.~R.}\ \bibnamefont {Akhmerov}}, \bibinfo
  {author} {\bibfnamefont {J.}~\bibnamefont {Tworzyd\l{}o}}, \ and\ \bibinfo
  {author} {\bibfnamefont {C.~W.~J.}\ \bibnamefont {Beenakker}},\ }\href
  {\doibase 10.1103/PhysRevLett.103.196805} {\bibfield  {journal} {\bibinfo
  {journal} {Phys. Rev. Lett.}\ }\textbf {\bibinfo {volume} {103}},\ \bibinfo
  {pages} {196805} (\bibinfo {year} {2009})}\BibitemShut {NoStop}%
\bibitem [{\citenamefont {Guo}\ \emph {et~al.}(2010)\citenamefont {Guo},
  \citenamefont {Rosenberg}, \citenamefont {Refael},\ and\ \citenamefont
  {Franz}}]{TAIin3D}%
  \BibitemOpen
  \bibfield  {author} {\bibinfo {author} {\bibfnamefont {H.-M.}\ \bibnamefont
  {Guo}}, \bibinfo {author} {\bibfnamefont {G.}~\bibnamefont {Rosenberg}},
  \bibinfo {author} {\bibfnamefont {G.}~\bibnamefont {Refael}}, \ and\ \bibinfo
  {author} {\bibfnamefont {M.}~\bibnamefont {Franz}},\ }\href {\doibase
  10.1103/PhysRevLett.105.216601} {\bibfield  {journal} {\bibinfo  {journal}
  {Phys. Rev. Lett.}\ }\textbf {\bibinfo {volume} {105}},\ \bibinfo {pages}
  {216601} (\bibinfo {year} {2010})}\BibitemShut {NoStop}%
\bibitem [{\citenamefont {Altland}\ \emph {et~al.}(2014)\citenamefont
  {Altland}, \citenamefont {Bagrets}, \citenamefont {Fritz}, \citenamefont
  {Kamenev},\ and\ \citenamefont {Schmiedt}}]{AltlandCrit}%
  \BibitemOpen
  \bibfield  {author} {\bibinfo {author} {\bibfnamefont {A.}~\bibnamefont
  {Altland}}, \bibinfo {author} {\bibfnamefont {D.}~\bibnamefont {Bagrets}},
  \bibinfo {author} {\bibfnamefont {L.}~\bibnamefont {Fritz}}, \bibinfo
  {author} {\bibfnamefont {A.}~\bibnamefont {Kamenev}}, \ and\ \bibinfo
  {author} {\bibfnamefont {H.}~\bibnamefont {Schmiedt}},\ }\href {\doibase
  10.1103/PhysRevLett.112.206602} {\bibfield  {journal} {\bibinfo  {journal}
  {Phys. Rev. Lett.}\ }\textbf {\bibinfo {volume} {112}},\ \bibinfo {pages}
  {206602} (\bibinfo {year} {2014})}\BibitemShut {NoStop}%
\bibitem [{\citenamefont {Titum}\ \emph {et~al.}(2015)\citenamefont {Titum},
  \citenamefont {Lindner}, \citenamefont {Rechtsman},\ and\ \citenamefont
  {Refael}}]{Titum-2015-DisFloquetTI}%
  \BibitemOpen
  \bibfield  {author} {\bibinfo {author} {\bibfnamefont {P.}~\bibnamefont
  {Titum}}, \bibinfo {author} {\bibfnamefont {N.~H.}\ \bibnamefont {Lindner}},
  \bibinfo {author} {\bibfnamefont {M.~C.}\ \bibnamefont {Rechtsman}}, \ and\
  \bibinfo {author} {\bibfnamefont {G.}~\bibnamefont {Refael}},\ }\href
  {\doibase 10.1103/PhysRevLett.114.056801} {\bibfield  {journal} {\bibinfo
  {journal} {Phys. Rev. Lett.}\ }\textbf {\bibinfo {volume} {114}},\ \bibinfo
  {pages} {056801} (\bibinfo {year} {2015})}\BibitemShut {NoStop}%
\bibitem [{\citenamefont {Liu}\ \emph {et~al.}(2017)\citenamefont {Liu},
  \citenamefont {Gao}, \citenamefont {Yang},\ and\ \citenamefont
  {Zhang}}]{Liu-Disorder-Top-Photonic}%
  \BibitemOpen
  \bibfield  {author} {\bibinfo {author} {\bibfnamefont {C.}~\bibnamefont
  {Liu}}, \bibinfo {author} {\bibfnamefont {W.}~\bibnamefont {Gao}}, \bibinfo
  {author} {\bibfnamefont {B.}~\bibnamefont {Yang}}, \ and\ \bibinfo {author}
  {\bibfnamefont {S.}~\bibnamefont {Zhang}},\ }\href {\doibase
  10.1103/PhysRevLett.119.183901} {\bibfield  {journal} {\bibinfo  {journal}
  {Phys. Rev. Lett.}\ }\textbf {\bibinfo {volume} {119}},\ \bibinfo {pages}
  {183901} (\bibinfo {year} {2017})}\BibitemShut {NoStop}%
\bibitem [{\citenamefont {Su}\ \emph {et~al.}(1979)\citenamefont {Su},
  \citenamefont {Schrieffer},\ and\ \citenamefont {Heeger}}]{SSH-1979}%
  \BibitemOpen
  \bibfield  {author} {\bibinfo {author} {\bibfnamefont {W.~P.}\ \bibnamefont
  {Su}}, \bibinfo {author} {\bibfnamefont {J.~R.}\ \bibnamefont {Schrieffer}},
  \ and\ \bibinfo {author} {\bibfnamefont {A.~J.}\ \bibnamefont {Heeger}},\
  }\href {\doibase 10.1103/PhysRevLett.42.1698} {\bibfield  {journal} {\bibinfo
   {journal} {Phys. Rev. Lett.}\ }\textbf {\bibinfo {volume} {42}},\ \bibinfo
  {pages} {1698} (\bibinfo {year} {1979})}\BibitemShut {NoStop}%
\bibitem [{\citenamefont {Mondragon-Shem}\ \emph {et~al.}(2014)\citenamefont
  {Mondragon-Shem}, \citenamefont {Hughes}, \citenamefont {Song},\ and\
  \citenamefont {Prodan}}]{DragonAIII}%
  \BibitemOpen
  \bibfield  {author} {\bibinfo {author} {\bibfnamefont {I.}~\bibnamefont
  {Mondragon-Shem}}, \bibinfo {author} {\bibfnamefont {T.~L.}\ \bibnamefont
  {Hughes}}, \bibinfo {author} {\bibfnamefont {J.}~\bibnamefont {Song}}, \ and\
  \bibinfo {author} {\bibfnamefont {E.}~\bibnamefont {Prodan}},\ }\href
  {\doibase 10.1103/PhysRevLett.113.046802} {\bibfield  {journal} {\bibinfo
  {journal} {Phys. Rev. Lett.}\ }\textbf {\bibinfo {volume} {113}},\ \bibinfo
  {pages} {046802} (\bibinfo {year} {2014})}\BibitemShut {NoStop}%
\bibitem [{\citenamefont {Meier}\ \emph
  {et~al.}(2016{\natexlab{a}})\citenamefont {Meier}, \citenamefont {An},\ and\
  \citenamefont {Gadway}}]{MSL-expt}%
  \BibitemOpen
  \bibfield  {author} {\bibinfo {author} {\bibfnamefont {E.~J.}\ \bibnamefont
  {Meier}}, \bibinfo {author} {\bibfnamefont {F.~A.}\ \bibnamefont {An}}, \
  and\ \bibinfo {author} {\bibfnamefont {B.}~\bibnamefont {Gadway}},\ }\href
  {\doibase 10.1103/PhysRevA.93.051602} {\bibfield  {journal} {\bibinfo
  {journal} {Phys. Rev. A}\ }\textbf {\bibinfo {volume} {93}},\ \bibinfo
  {pages} {051602} (\bibinfo {year} {2016}{\natexlab{a}})}\BibitemShut
  {NoStop}%
\bibitem [{\citenamefont {Lohse}\ \emph {et~al.}(2016)\citenamefont {Lohse},
  \citenamefont {Schweizer}, \citenamefont {Zilberberg}, \citenamefont
  {Aidelsburger},\ and\ \citenamefont {Bloch}}]{Lohse-pumping}%
  \BibitemOpen
  \bibfield  {author} {\bibinfo {author} {\bibfnamefont {M.}~\bibnamefont
  {Lohse}}, \bibinfo {author} {\bibfnamefont {C.}~\bibnamefont {Schweizer}},
  \bibinfo {author} {\bibfnamefont {O.}~\bibnamefont {Zilberberg}}, \bibinfo
  {author} {\bibfnamefont {M.}~\bibnamefont {Aidelsburger}}, \ and\ \bibinfo
  {author} {\bibfnamefont {I.}~\bibnamefont {Bloch}},\ }\href {\doibase
  10.1038/nphys3584} {\bibfield  {journal} {\bibinfo  {journal} {Nat. Phys.}\
  }\textbf {\bibinfo {volume} {12}},\ \bibinfo {pages} {350} (\bibinfo {year}
  {2016})}\BibitemShut {NoStop}%
\bibitem [{\citenamefont {Nakajima}\ \emph {et~al.}(2016)\citenamefont
  {Nakajima}, \citenamefont {Tomita}, \citenamefont {Taie}, \citenamefont
  {Ichinose}, \citenamefont {Ozawa}, \citenamefont {Wang}, \citenamefont
  {Troyer},\ and\ \citenamefont {Takahashi}}]{Nakajima-pumping}%
  \BibitemOpen
  \bibfield  {author} {\bibinfo {author} {\bibfnamefont {S.}~\bibnamefont
  {Nakajima}}, \bibinfo {author} {\bibfnamefont {T.}~\bibnamefont {Tomita}},
  \bibinfo {author} {\bibfnamefont {S.}~\bibnamefont {Taie}}, \bibinfo {author}
  {\bibfnamefont {T.}~\bibnamefont {Ichinose}}, \bibinfo {author}
  {\bibfnamefont {H.}~\bibnamefont {Ozawa}}, \bibinfo {author} {\bibfnamefont
  {L.}~\bibnamefont {Wang}}, \bibinfo {author} {\bibfnamefont {M.}~\bibnamefont
  {Troyer}}, \ and\ \bibinfo {author} {\bibfnamefont {Y.}~\bibnamefont
  {Takahashi}},\ }\href {\doibase 10.1038/nphys3622} {\bibfield  {journal}
  {\bibinfo  {journal} {Nat. Phys.}\ }\textbf {\bibinfo {volume} {12}},\
  \bibinfo {pages} {296} (\bibinfo {year} {2016})}\BibitemShut {NoStop}%
\bibitem [{\citenamefont {Velasco}\ and\ \citenamefont
  {Paredes}(2017)}]{Velasco-2017-AIII}%
  \BibitemOpen
  \bibfield  {author} {\bibinfo {author} {\bibfnamefont {C.~G.}\ \bibnamefont
  {Velasco}}\ and\ \bibinfo {author} {\bibfnamefont {B.}~\bibnamefont
  {Paredes}},\ }\href {\doibase 10.1103/PhysRevLett.119.115301} {\bibfield
  {journal} {\bibinfo  {journal} {Phys. Rev. Lett.}\ }\textbf {\bibinfo
  {volume} {119}},\ \bibinfo {pages} {115301} (\bibinfo {year}
  {2017})}\BibitemShut {NoStop}%
\bibitem [{\citenamefont {Billy}\ \emph {et~al.}(2008)\citenamefont {Billy},
  \citenamefont {Josse}, \citenamefont {Zuo}, \citenamefont {Bernard},
  \citenamefont {Hambrecht}, \citenamefont {Lugan}, \citenamefont
  {Cl\'{e}ment}, \citenamefont {Sanchez-Palencia}, \citenamefont {Bouyer},\
  and\ \citenamefont {Aspect}}]{Billy-AndersonLocalization-2008}%
  \BibitemOpen
  \bibfield  {author} {\bibinfo {author} {\bibfnamefont {J.}~\bibnamefont
  {Billy}}, \bibinfo {author} {\bibfnamefont {V.}~\bibnamefont {Josse}},
  \bibinfo {author} {\bibfnamefont {Z.}~\bibnamefont {Zuo}}, \bibinfo {author}
  {\bibfnamefont {A.}~\bibnamefont {Bernard}}, \bibinfo {author} {\bibfnamefont
  {B.}~\bibnamefont {Hambrecht}}, \bibinfo {author} {\bibfnamefont
  {P.}~\bibnamefont {Lugan}}, \bibinfo {author} {\bibfnamefont
  {D.}~\bibnamefont {Cl\'{e}ment}}, \bibinfo {author} {\bibfnamefont
  {L.}~\bibnamefont {Sanchez-Palencia}}, \bibinfo {author} {\bibfnamefont
  {P.}~\bibnamefont {Bouyer}}, \ and\ \bibinfo {author} {\bibfnamefont
  {A.}~\bibnamefont {Aspect}},\ }\href {\doibase 10.1038/nature07000}
  {\bibfield  {journal} {\bibinfo  {journal} {Nature}\ }\textbf {\bibinfo
  {volume} {453}},\ \bibinfo {pages} {891} (\bibinfo {year}
  {2008})}\BibitemShut {NoStop}%
\bibitem [{\citenamefont {Roati}\ \emph {et~al.}(2008)\citenamefont {Roati},
  \citenamefont {D'Errico}, \citenamefont {Fallani}, \citenamefont {Fattori},
  \citenamefont {Fort}, \citenamefont {Zaccanti}, \citenamefont {Modugno},
  \citenamefont {Modugno},\ and\ \citenamefont
  {Inguscio}}]{Roati-AndersonLocalization-2008}%
  \BibitemOpen
  \bibfield  {author} {\bibinfo {author} {\bibfnamefont {G.}~\bibnamefont
  {Roati}}, \bibinfo {author} {\bibfnamefont {C.}~\bibnamefont {D'Errico}},
  \bibinfo {author} {\bibfnamefont {L.}~\bibnamefont {Fallani}}, \bibinfo
  {author} {\bibfnamefont {M.}~\bibnamefont {Fattori}}, \bibinfo {author}
  {\bibfnamefont {C.}~\bibnamefont {Fort}}, \bibinfo {author} {\bibfnamefont
  {M.}~\bibnamefont {Zaccanti}}, \bibinfo {author} {\bibfnamefont
  {G.}~\bibnamefont {Modugno}}, \bibinfo {author} {\bibfnamefont
  {M.}~\bibnamefont {Modugno}}, \ and\ \bibinfo {author} {\bibfnamefont
  {M.}~\bibnamefont {Inguscio}},\ }\href {\doibase 10.1038/nature07071}
  {\bibfield  {journal} {\bibinfo  {journal} {Nature}\ }\textbf {\bibinfo
  {volume} {453}},\ \bibinfo {pages} {895} (\bibinfo {year}
  {2008})}\BibitemShut {NoStop}%
\bibitem [{\citenamefont {Meier}\ \emph
  {et~al.}(2016{\natexlab{b}})\citenamefont {Meier}, \citenamefont {An},\ and\
  \citenamefont {Gadway}}]{MSL-edge}%
  \BibitemOpen
  \bibfield  {author} {\bibinfo {author} {\bibfnamefont {E.~J.}\ \bibnamefont
  {Meier}}, \bibinfo {author} {\bibfnamefont {F.~A.}\ \bibnamefont {An}}, \
  and\ \bibinfo {author} {\bibfnamefont {B.}~\bibnamefont {Gadway}},\ }\href
  {\doibase 10.1038/ncomms13986} {\bibfield  {journal} {\bibinfo  {journal}
  {Nat. Commun.}\ }\textbf {\bibinfo {volume} {7}},\ \bibinfo {pages} {13986}
  (\bibinfo {year} {2016}{\natexlab{b}})}\BibitemShut {NoStop}%
\bibitem [{\citenamefont {Cardano}\ \emph {et~al.}(2017)\citenamefont
  {Cardano}, \citenamefont {D'Errico}, \citenamefont {Dauphin}, \citenamefont
  {Maffei}, \citenamefont {Piccirillo}, \citenamefont {de~Lisio}, \citenamefont
  {De~Filippis}, \citenamefont {Cataudella}, \citenamefont {Santamato},
  \citenamefont {Marrucci}, \citenamefont {Lewenstein},\ and\ \citenamefont
  {Massignan}}]{photonicCD}%
  \BibitemOpen
  \bibfield  {author} {\bibinfo {author} {\bibfnamefont {F.}~\bibnamefont
  {Cardano}}, \bibinfo {author} {\bibfnamefont {A.}~\bibnamefont {D'Errico}},
  \bibinfo {author} {\bibfnamefont {A.}~\bibnamefont {Dauphin}}, \bibinfo
  {author} {\bibfnamefont {M.}~\bibnamefont {Maffei}}, \bibinfo {author}
  {\bibfnamefont {B.}~\bibnamefont {Piccirillo}}, \bibinfo {author}
  {\bibfnamefont {C.}~\bibnamefont {de~Lisio}}, \bibinfo {author}
  {\bibfnamefont {G.}~\bibnamefont {De~Filippis}}, \bibinfo {author}
  {\bibfnamefont {V.}~\bibnamefont {Cataudella}}, \bibinfo {author}
  {\bibfnamefont {E.}~\bibnamefont {Santamato}}, \bibinfo {author}
  {\bibfnamefont {L.}~\bibnamefont {Marrucci}}, \bibinfo {author}
  {\bibfnamefont {M.}~\bibnamefont {Lewenstein}}, \ and\ \bibinfo {author}
  {\bibfnamefont {P.}~\bibnamefont {Massignan}},\ }\href {\doibase
  10.1038/ncomms15516} {\bibfield  {journal} {\bibinfo  {journal} {Nat.
  Commun.}\ }\textbf {\bibinfo {volume} {8}},\ \bibinfo {pages} {15516}
  (\bibinfo {year} {2017})}\BibitemShut {NoStop}%
\bibitem [{\citenamefont {Maffei}\ \emph {et~al.}(2018)\citenamefont {Maffei},
  \citenamefont {Dauphin}, \citenamefont {Cardano}, \citenamefont
  {Lewenstein},\ and\ \citenamefont {Massignan}}]{MariaNJP}%
  \BibitemOpen
  \bibfield  {author} {\bibinfo {author} {\bibfnamefont {M.}~\bibnamefont
  {Maffei}}, \bibinfo {author} {\bibfnamefont {A.}~\bibnamefont {Dauphin}},
  \bibinfo {author} {\bibfnamefont {F.}~\bibnamefont {Cardano}}, \bibinfo
  {author} {\bibfnamefont {M.}~\bibnamefont {Lewenstein}}, \ and\ \bibinfo
  {author} {\bibfnamefont {P.}~\bibnamefont {Massignan}},\ }\href
  {http://stacks.iop.org/1367-2630/20/i=1/a=013023} {\bibfield  {journal}
  {\bibinfo  {journal} {New J. Phys.}\ }\textbf {\bibinfo {volume} {20}},\
  \bibinfo {pages} {013023} (\bibinfo {year} {2018})}\BibitemShut {NoStop}%
\bibitem [{Sup()}]{Supplement}%
  \BibitemOpen
  \href@noop {} {}\bibinfo {note} {See the supplementary materials accompanying
  this document for more information on the relationship between the winding
  number topological invariant and the mean chiral displacement.}\BibitemShut
  {Stop}%
\bibitem [{\citenamefont {Altland}\ \emph {et~al.}(2015)\citenamefont
  {Altland}, \citenamefont {Bagrets},\ and\ \citenamefont
  {Kamenev}}]{Altland-PRB2015}%
  \BibitemOpen
  \bibfield  {author} {\bibinfo {author} {\bibfnamefont {A.}~\bibnamefont
  {Altland}}, \bibinfo {author} {\bibfnamefont {D.}~\bibnamefont {Bagrets}}, \
  and\ \bibinfo {author} {\bibfnamefont {A.}~\bibnamefont {Kamenev}},\ }\href
  {\doibase 10.1103/PhysRevB.91.085429} {\bibfield  {journal} {\bibinfo
  {journal} {Phys. Rev. B}\ }\textbf {\bibinfo {volume} {91}},\ \bibinfo
  {pages} {085429} (\bibinfo {year} {2015})}\BibitemShut {NoStop}%
\bibitem [{\citenamefont {An}\ \emph {et~al.}(2018)\citenamefont {An},
  \citenamefont {Meier}, \citenamefont {Ang'ong'a},\ and\ \citenamefont
  {Gadway}}]{MSL-interactions}%
  \BibitemOpen
  \bibfield  {author} {\bibinfo {author} {\bibfnamefont {F.~A.}\ \bibnamefont
  {An}}, \bibinfo {author} {\bibfnamefont {E.~J.}\ \bibnamefont {Meier}},
  \bibinfo {author} {\bibfnamefont {J.}~\bibnamefont {Ang'ong'a}}, \ and\
  \bibinfo {author} {\bibfnamefont {B.}~\bibnamefont {Gadway}},\ }\href
  {\doibase 10.1103/PhysRevLett.120.040407} {\bibfield  {journal} {\bibinfo
  {journal} {Phys. Rev. Lett.}\ }\textbf {\bibinfo {volume} {120}},\ \bibinfo
  {pages} {040407} (\bibinfo {year} {2018})}\BibitemShut {NoStop}%
\end{thebibliography}

%

%
%

\end{document}


\title{Supplementary Materials for: Observation of the topological Anderson insulator in disordered atomic wires}

\author{Eric J. Meier}
\affUIUC

\author{Fangzhao Alex An}
\affUIUC

\author{Alexandre Dauphin}
\affICFO

\author{Maria Maffei}
\affICFO
\affMariatwo

\author{Pietro Massignan}
\email{pietro.massignan@upc.edu}
\affICFO
\affPolytech

\author{Taylor L. Hughes}
\email{hughest@illinois.edu}
\affUIUC

\author{Bryce Gadway}
\email{bgadway@illinois.edu}
\affUIUC

\date{\today}

\renewcommand\thefigure{S\arabic{figure}}
\renewcommand\theequation{S\arabic{equation}}

\maketitle

\section{Experimental setup}

All the experiments shown in the main text begin with Bose--Einstein condensates (BECs) of $^{87}$Rb with roughly $10^5$ atoms. The BECs are optically trapped by two crossed 1064~nm laser beams, with most of the trapping power in just a single beam and the other providing weak confinement. This arrangement results in a weak harmonic trapping along the propagation axis of the high-power beam (with a harmonic trapping frequency of roughly 10~Hz) and tighter trapping (with a harmonic frequency of roughly 130 Hz) along the other two axes.

The lattice is created by passing the high-power trapping beam through a series of acousto-optic modulators which turn the single frequency beam into a beam containing many slightly detuned frequency components. This multi-frequency beam is then directed to counter-propagate with itself at the location of the atoms to drive the two-photon Bragg transitions linking sites in the momentum-space lattice as shown in Fig.~\pref{fig1}{a}. Since the different synthetic lattice sites have different momenta, we can image the atomic population at a site-resolved level by performing time-of-flight imaging, where the time-of-flight period allows the atoms at different lattice sites to separate in real space.

\section{Spectroscopic engineering of effective tight-binding models}

\begin{figure}%
\includegraphics[width=500pt]{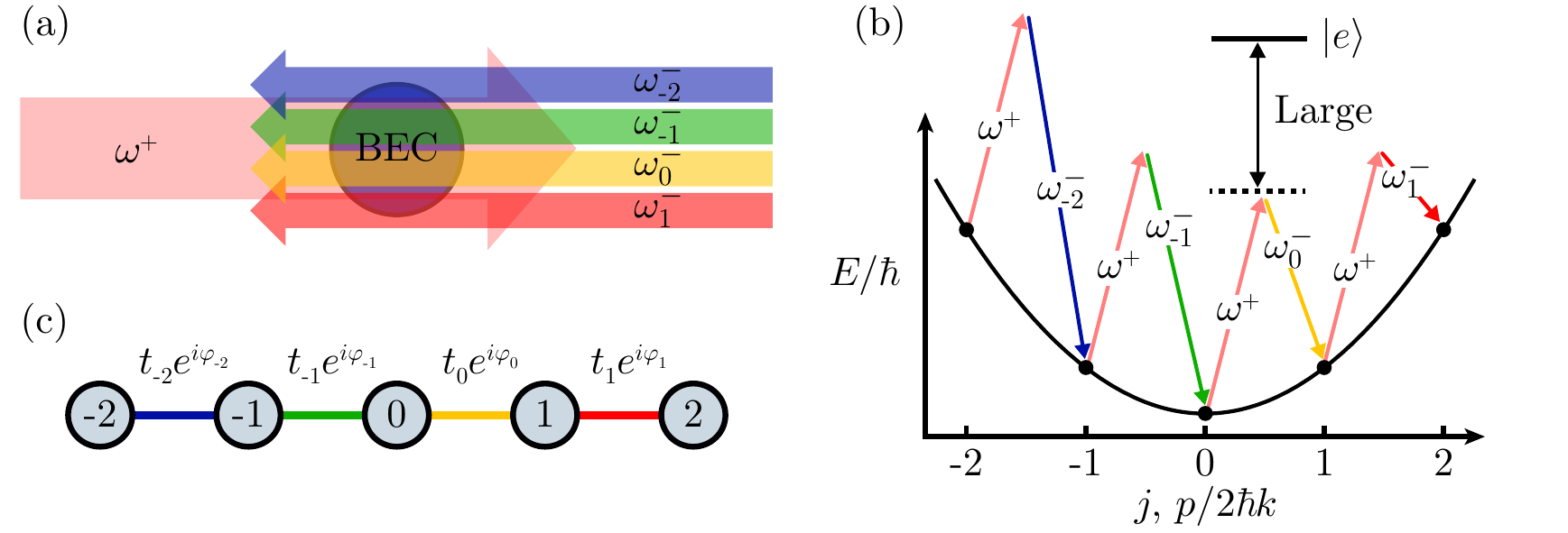}%
\caption{Experimental setup.
\textbf{(a)} Cartoon depiction of the momentum-space lattice lasers incident on the Bose--Einstein condensate (BEC).
\textbf{(b)} Free-particle dispersion relation of individual atoms showing the available momentum states as points along the parabola and the lattice lasers from \textbf{(a)} linking the states.
\textbf{(c)} The resulting momentum space lattice formed by the tunnel couplings shown in \textbf{(a)} and \textbf{(b)}. This figure is adapted from Ref.~\cite{MSL-disorderdynamics}.
}%
\label{fig1}%
\end{figure}

A discrete set of states $\psi_j$ with momenta $p_j = 2j\hbar k$ are determined by the lattice laser wavevector $k$ ($k = 2\pi/\lambda$ with laser wavelength $\lambda$). These states may be populated from a zero-momentum condensate through the stimulated exchange of photons between the two counter-propagating laser fields as shown in Fig.~\pref{fig1}{a}. A unique energy difference and Bragg transition frequency $\omega^{res}_j = (2j+1)4 E_R/\hbar$ (with $E_R = \hbar^2 k^2 / 2M$ the recoil energy and $M$ the mass of the atoms) between each pair of neighboring states $\psi_j$ and $\psi_{j+1}$ is defined by the quadratic dispersion of the atoms as shown in Fig.~\pref{fig1}{b}. By writing multiple frequency components onto the counter-propagating laser field, each of which addresses a unique state-to-state Bragg transition, we achieve spectrally-resolved control over all effective lattice parameters at the individual link level.

More formally, at the single-particle level the momentum-space evolution of the atoms is governed by the Hamiltonian $H = H_0 + V(t)$. Here, $H_0 = \sum_j E_j \ket{\psi_j}\bra{\psi_j}$ describes the kinetic energies $E_j =4 j^2 E_R$ of the $\psi_j$ states and $V(t) = \sum_j (\chi (t) \ket{\psi_{j+1}}\bra{\psi_j} + \chi^* (t) \ket{\psi_{j}}\bra{\psi_{j+1}})$ describes the interaction of atoms with the counter-propagating laser field. The common (for all $j$) off-diagonal coupling constant is defined by $\chi(t) = \sum_l \hbar\widetilde{\Omega}_l e^{i \varphi_l} e^{-i \widetilde{\omega}_l t}$, and relates to changes of momenta by $+2\hbar k$ via virtual absorption of a photon from the right-traveling beam and stimulated emission into one of the left-traveling fields with index $l$ (with right/left directions as depicted in Fig.~\pref{fig1}{a}). Here, the two-photon coupling strengths are $\widetilde{\Omega}_l = \Omega^-_l \Omega^+ / 2\Delta$ in terms of the single-photon detuning $\Delta$ from atomic resonance (ground $\ket{g}$ to excited $\ket{e}$ state transition as shown in Fig.~\pref{fig1}{b}) and the single-photon Rabi couplings (assumed to be real-valued), with $\Omega^+$ relating to the right-traveling field and $\Omega^-_l$ to the $l^{\mathrm{th}}$ frequency component of the left-traveling field. Similarly, the two-photon phase shift is related to the phase difference between the interfering fields as $\varphi_l = \phi^+ - \phi^{-}_l$, and the two-photon frequency is given by the frequency difference between the interfering fields as $\widetilde{\omega}_l = \omega^{-}_l - \omega^+$. We can move to the interaction picture to remove the diagonal kinetic energy terms, leaving only $H^{\mathrm{int}}=V_I(t)=e^{iH_0t/\hbar}V(t)e^{-iH_0t/\hbar}$, with $V_I(t)=\sum_j(\widetilde{\chi}_j(t)\ket{\psi_{j+1}}\bra{\psi_j}+\widetilde{\chi}_j^*(t)\ket{\psi_{j}}\bra{\psi_{j+1}})$ and $\widetilde{\chi}_j(t)=\chi(t)e^{i(E_{j+1}-E_j)t/\hbar}=\chi(t)e^{i\omega^{res}_jt}$.

We achieve our goal of uniquely controlling all nearest-neighbor couplings at the single-link level by associating each frequency component with a unique Bragg resonance. Specifically, we define $\widetilde{\omega}_l = \omega^{res}_l$, such that $\chi(t) = \sum_l \hbar\widetilde{\Omega}_l e^{i \varphi_l} e^{i\omega^{res}_l t}$. In the weak-driving limit ($\hbar \widetilde{\Omega}_j \ll 8 E_R \ \forall \ j$), each frequency component contributes to the coupling of only two particular momentum states $\psi_j$ and $\psi_{j+1}$. Ignoring all rapidly oscillating terms, we arrive at an approximate description with controlled, nearest-neighbor couplings $\expval{j+1}{V_I(t)}{j} \approx \hbar\widetilde{\Omega}_j e^{i \varphi_j}$. This brings us in final form to a highly tunable single-particle Hamiltonian
\begin{equation}
H_{\textrm{eff}} \approx \sum_j t_j ( e^{i\varphi_j} \ket{\widetilde{\psi}_{j+1}}\bra{\widetilde{\psi}_j} + \textrm{h.c.}),
\label{ham}
\end{equation}
where arbitrary control over all tunneling amplitudes $t_j \equiv \hbar \widetilde{\Omega}_j$ and tunneling phases $\varphi_j$ is enabled in a link-dependent fashion through control of the multi-frequency global addressing field. This gives us the ability to control $t_j$ and $\varphi_j$ not only locally but also time-dependently. Here, however, the only time dependence in these parameters is that they are turned on suddenly in a step-wise fashion.

In addition to the single-particle dynamics driven by atom-light interactions, this cold atom system can also have important contributions from real-space contact interactions between the atoms. In momentum space, for a one-dimensional system, atomic scattering is predominantly momentum mode-conserving due to energy conservation and the quadratic free-particle dispersion. This leads to self- and cross-phase modulation terms between the various momentum modes, at the scale of the condensate mean-field energy $U=4\pi\hbar^2 a \rho_N$ (with $a$ the scattering length and $\rho_N$ the uniform atomic number density in real space)~\cite{Trippenbach4wave}. Due to bosonic statistics, there is an added exchange energy between atoms scattering in distinguishable momentum states. This results in cross-phase modulation terms that are twice as large as the self-phase modulation terms. For repulsive real-space interactions, as in the case of $^{87}$Rb, this added long-range (in momentum-space) repulsion can be recast as an effective attraction between atoms in the same momentum mode. We can approximately capture the influence of interactions in our system by the time-dependent Gross-Pitaevskii equation
\begin{equation}
i \hbar \dot{\psi}_j=t_je^{-i\varphi_j}\psi_{j+1}+t_{j-1}e^{i\varphi_{j-1}}\psi_{j-1}+U\left(2-|\psi_j|^2\right)\psi_j,
\end{equation}
where $\psi_j$ are the normalized momentum-mode amplitudes and the tunneling elements relate to those of $H_\textrm{eff}$. While these interactions may enable us to study interacting topological quantum fluids in the future, this experiment was designed to minimize the effects due to interactions. As such we find no evidence of significant effects due to interactions in the data presented here or in the main text. For more information on the role of interactions in the momentum-space lattice see Ref.~\cite{MSL-interactions}.

\section{Evidence for the TAI phase in BDI wires}

In addition to the study in the main text that explores the influence of tunneling disorder on AIII-class wires, here we study and provide evidence for a topological Anderson insulating (TAI) phase in BDI-class wires. The case of disordered AIII-class wires in the main text utilized disorder that was only applied to the intra-cell tunneling elements, similar to the theory study of Ref.~\cite{Altland-PRB2015}. However, here, for the BDI wires, we consider the same ratio of disorder strengths on the intra- and inter-cell terms as was examined in theory in Ref.~\cite{DragonAIII} and experimentally for Fig.~2 of the main text, i.e. $W \equiv W_2 = 2W_1$. The consequence of this disorder arrangement is that the region of the topological phase diagram relating to the TAI is somewhat smaller than for the scenario in which $W_1 = 0$. Because of finite-time broadening, the measurement of $\langle \bar{\C} \rangle$ takes a smaller maximum value in this case as compared to the case of fully asymmetric ($W_1 = 0$) disorder, as the TAI phase is sandwiched between two trivial regions (for constant tunneling imbalance $m$ as a function of disorder strength $W$).

\begin{figure*}
\includegraphics[width=500pt]{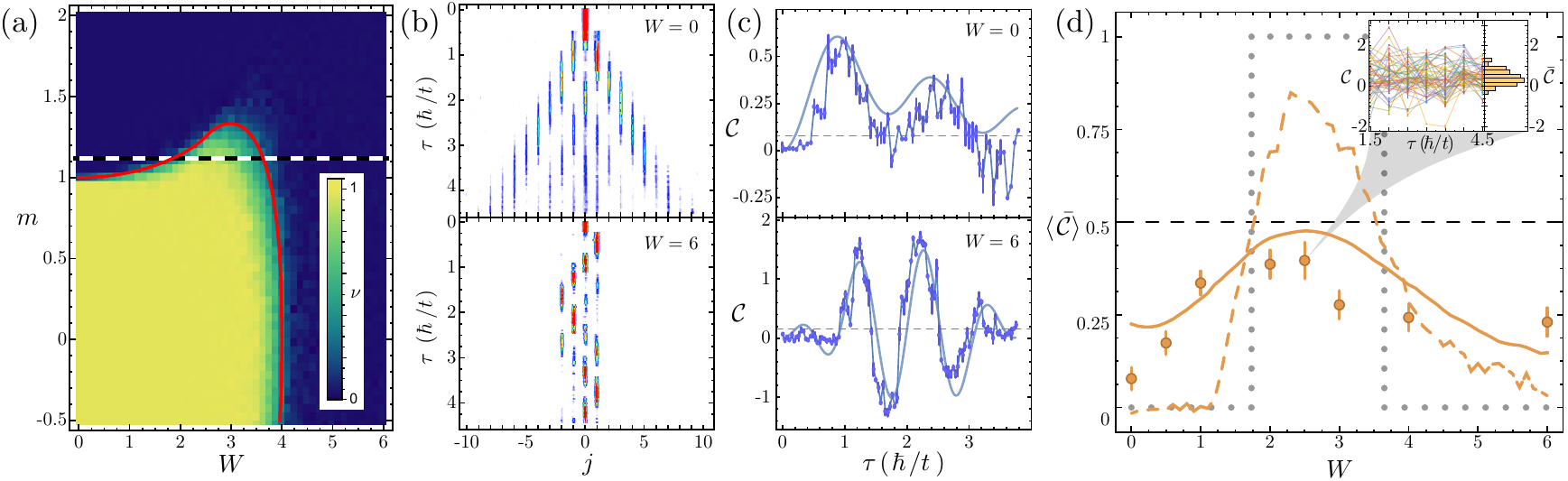}
\caption{Probing the topological Anderson insulating phase in BDI class wires.
\textbf{(a)} Topological phase diagram of the BDI wire model, showing the computed winding number (inset color scale) as a function of disorder strength $W$ and tunneling ratio $m$ with tunneling disorder strengths $W\equiv W_2 = 2W_1$. The dashed horizontal line at $m = 1.12$ indicates the region explored experimentally. The solid red curve indicates the critical boundary (i.e. the set of points where the localization length diverges for an infinite chain).
\textbf{(b)} Integrated absorption images of the bulk dynamics following a sudden quench of the tunnel couplings, for both zero disorder ($W = 0$) and strong disorder ($W = 6$, for a single configuration of the disorder).
\textbf{(c)} Dynamics of $\C$ for the same data as in \textbf{(b)}. The solid blue curves are numerical simulations with no free parameters. The dashed gray horizontal lines denote $\bar{\C}$ for the data.
\textbf{(d)} $\langle \bar{\C} \rangle$ as a function of $W$ for $m=1.12(2)$. The data are averaged over $50$ disorder configurations and times in the range 1.5~$\hbar/t$ to $4.5~\hbar/t$ in steps of $0.5~\hbar/t$. The solid gold line represents a numerical simulation based on the exact experimental time and disorder averaging. The dashed gold line is based on the same simulation as the solid gold line, but sampled to much longer times ($\tau$ = 1,000~$\hbar/t$ in a system with 250 unit cells).
The dotted grey curve shows the topological index in the thermodynamic limit~\cite{DragonAIII}, which takes a value of 0.5 at the critical points, as indicated by the horizontal dashed line.
The inset shows $\C$ as a function of time for all 50 disorder configurations (for $W = 2.5$) and a histogram of the corresponding $\bar{\C}$ values.
All error bars in \textbf{(c)} and \textbf{(d)} denote one standard error of the mean.
}
\label{fig2}
\end{figure*}

The topological phase diagram for the BDI wires as a function of $m$ and $W$, equivalent to that of Fig.~2(a) of the BDI case in the main text, is reproduced here in Fig.~\pref{fig2}{a}. Here, we measure $\langle\bar{\C}\rangle$ along the line $m = 1.12(2)$, by observing the bulk density response to a quench of the tunneling terms, as in the main text. The bulk response for zero disorder ($W = 0$) and large disorder ($W = 6$, for a single configuration of the disorder) are shown in Fig.~\pref{fig2}{b} and the corresponding $\C$ measurement is shown in Fig.~\pref{fig2}{c}, where we find good agreement with a zero-free-parameter numerical simulation (blue lines).

The measurements of $\langle \bar{\C} \rangle$ as a function of $W$ are shown in Fig.~\pref{fig2}{d}, taken for 50 disorder configurations and for times ranging from 0.5~$\hbar/t$ to 4.5~$\hbar/t$, in steps of 0.5~$\hbar/t$ for $t/\hbar\approx 2\pi\times$600~Hz. As in the case of AIII-class wires, here we observe that the addition of weak to moderate disorder leads to an increase in $\langle \bar{\C} \rangle$, while strong disorder causes $\langle \bar{\C} \rangle$ to decay again. The data is in agreement with a numerical simulation (gold solid line in Fig.~\pref{fig2}{d}) based on the experimental time sampling. The sharp rise and fall of $\langle \bar{\C} \rangle$ found in simulations extended out to much longer times (dashed gold line, for 1,000 tunneling times) suggest that we are observing successive topological phase transitions that are broadened due to finite time sampling. To note, the maximum $\langle \bar{\C} \rangle$ measurement and even the equivalent-time theory predictions are slightly less than 0.5 in this case, due to the different ratio of disorder strengths applied to intra- and inter-cell tunnelings as compared to the AIII data of Fig.~3 of the main text.

\section{Tunneling calibrations}

To calibrate the tunneling strength $t$ in these experiments, we expose the condensate to a single frequency component linking site $j=0$ to site $j=1$. We then observe Rabi oscillations over several tunneling times. The measured Rabi oscillation frequency allows us to extract the coupling strength $t$ between the two sites. By ensuring that this calibration laser field has the exact same strength as the experimental laser field we are able to calibrate the experimental tunneling strength.

\section{Comparison of full and approximated numerical simulations}

\begin{figure*}[h]
\includegraphics[width=486pt]{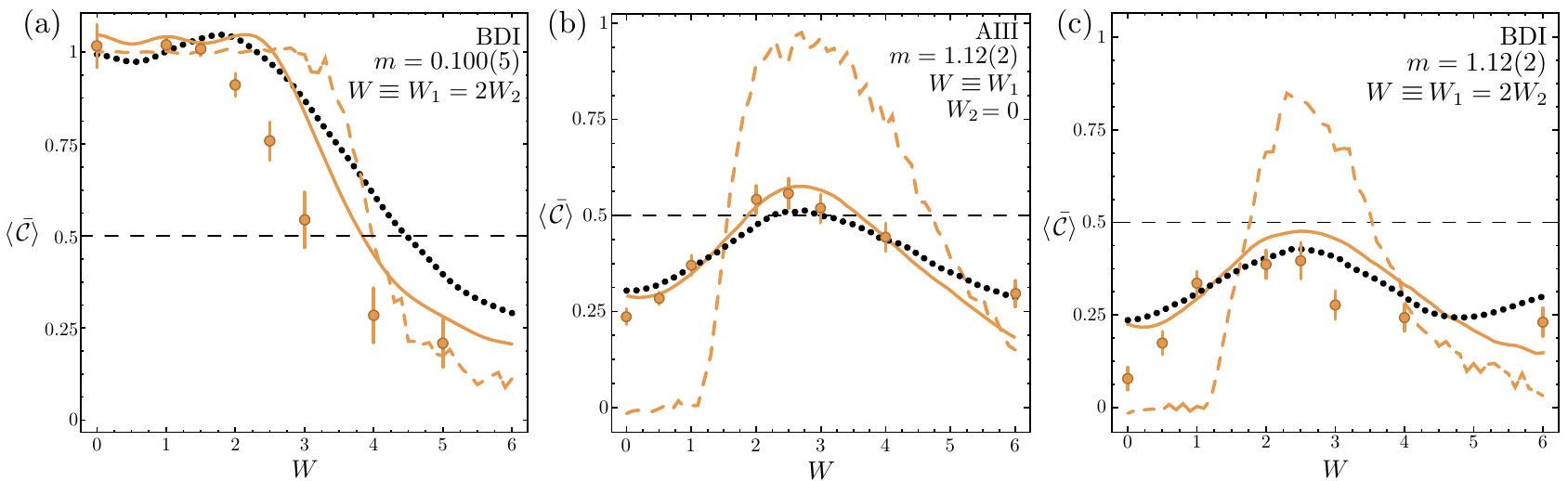}
\caption{Comparison of full and ideal simulations to data.
\textbf{(a)} $\langle \bar{\C}\rangle$ as a function of $W$ for a BDI-class wire with $W\equiv 2W_1=W_2$, $m=0.100(5)$, and $t/\hbar\approx 2\pi\times$1200~Hz. The dotted black line is a ``full'' simulation including off-resonant driving and interaction effects with $U= 2\pi \times 700$~Hz as measured through Bragg spectroscopy. These are compared to the ``ideal'' simulations presented in the main text for experimental system sizes and timescales (solid gold line) and for larger systems with 250 unit cells and timescales of 1,000~$\hbar/t$ (dashed gold line).
\textbf{(b)} $\langle \bar{\C}\rangle$ as a function of $W$ for a AIII-class wire with $W_1=0$ and $W\equiv W_2$, $m=1.12(2)$, and $t/\hbar\approx 2\pi\times$600~Hz. The dotted black line is a full simulation and the gold lines are ideal simulations as presented in the main text for small (solid line, experimental sizes and timescales) and large (dashed line, 250 unit cells and 1,000~$\hbar/t$) systems.
\textbf{(c)} $\langle \bar{\C}\rangle$ as a function of $W$ for a BDI-class wire with $W\equiv 2W_1=W_2$, $m=1.12(2)$, and $t/\hbar\approx 2\pi\times$600~Hz. The dotted black line is a full simulation and the gold lines are ideal simulations as presented in the main text for small (solid line, experimental sizes and timescales) and large (dashed line, 250 unit cells and 1,000~$\hbar/t$) systems.
All error bars denote one standard error of the mean.
}
\label{fig3}
\end{figure*}

While the data presented in Figs.~2~and~3 of the main text and Fig.~\ref{fig2} of the supplement agree very well with the ``ideal'' numerical simulations of Eq.~\eqref{ham}, we also performed simulations that more exactly represent our system, i.e. including effects due to off-resonant driving and interactions (assuming an interaction energy $U \approx 2\pi \times 700$~Hz based on Bragg spectroscopy). The results of these ``full'' simulations are compared to their respective data and ideal simulations in Fig.~\ref{fig3}. We find, in general, that the full simulation and ideal simulation both agree with the data and each other adequately such that the rotating wave approximation which gives the simplified form of Eq.~\eqref{ham} of the main text is reasonably justified.

\section{Winding number and mean chiral displacement for disordered 1D chiral models}

The topology of 1D chiral models in classes AIII and BDI is characterized by the winding number $\nu$. In translationally invariant systems the winding number is most simply calculated in momentum space using the Bloch wavefunctions of the energy bands. However, in the presence of disorder, where translational symmetry is broken, the winding number must be computed in real space as shown in Ref.~\cite{DragonAIII},  and references therein. Here we revisit the real-space formalism, and present an alternative method that can be used to extract the winding number in disordered systems even with open boundary conditions. Subsequently, we show that in the thermodynamic limit the winding number of the disordered fermion system of our model matches the expectation value of the chiral displacement operator. 

Given a generic Hamiltonian $H$, let us introduce its ``flat-band analogue'' $Q=P_+-P_-$, defined as the projector on the positive-energy eigenstates minus the projector on the negative-energy eigenstates.
If the original Hamiltonian is chiral, so is the Q-matrix, and it is therefore possible to write it as
\beq
Q=Q_{AB} + Q_{BA} = \Gamma_A Q \Gamma_B + \Gamma_B Q \Gamma_A,
\eeq
where $\Gamma_A, \Gamma_B$ are projectors onto the $A$ or $B$ sublattices respectively, and $\Gamma=\Gamma_A - \Gamma_B$ is the chiral operator.
The winding number is given by
\beq
\label{realspacewindingnumber}
\nu=\int_0^{2\pi}\frac{{\rm d}k}{2\pi}{\rm Tr}[(Q_{AB})^{-1}\,i\partial_k Q_{AB} ]= \mathcal{T}\{Q_{BA}[X,Q_{AB}]\}=\mathcal{T}\{Q_{BA}X Q_{AB} - Q_{BA}Q_{AB}X\},
\eeq
where ${\rm Tr}$ indicates a trace over the unit cell, and $\mathcal{T}$ indicates a ``trace per volume'' (i.e., per number of unit cells $N$). The expression in momentum space (here referring to the generalized quasi-momentum space of the considered tight-binding model, and not the physical momentum of the atoms) assumes periodic boundary conditions, since translation symmetry is a necessary ingredient for the momentum space version of the formula. In position space, one must be careful when evaluating this expression. The conventional (Hermitian) position operator is not a valid operator when periodic boundary conditions are assumed, and a naive evaluation of the expression can give spurious results. If one wants to maintain periodic boundary conditions then one must either carefully evaluate the matrix elements of the $X$ operator such that periodic boundary conditions are imposed, or instead calculate the Berry-Zak phase $\phi_B$ in real-space using the formulation in Ref.~\cite{resta1998}. For our system the latter method would be sufficient because the winding number only takes values $\nu=0, \pm 1,$ and it was proven that in general $\nu\mod 2=\phi_{B}/\pi\mod 2$~\cite{DragonAIII}. Hence, one could use the Berry-Zak phase to unambiguously determine which phases of our model are topological and which are trivial.

The real-space formulation of the winding number provides an accurate representation of the disordered phases and the location of the disorder-driven topological phase transition in our model. Operatively, however, a single determination of the winding within this formulation requires a measure over all possible eigenstates of the model, which is not trivial to perform. To access the winding experimentally, we proceed along a very different way, and directly measure the mean chiral displacement, a quantity which upon sufficient time- and disorder-average converges to the winding number. For clean systems, this convergence has already been proven analytically in Refs.~\cite{photonicCD,Marianjp}. As we will explain in the following, the identification works also in presence of disorder.

To extend the proof to disordered systems, let us start from the static case, where we can use the results of Ref.~\cite{DragonAIII}. At the end of that article it was shown that the ground state of our model, for any disorder strength, takes the form of a ``random-singlet" (or ``random-dimer") configuration, i.e., the many-body ground state is a Slater determinant of single-particle states that are equal-weight linear combinations of exactly two sites of the lattice, one on sublattice $A$ and one on $B$. Hence, we can write the aysmptotically exact wavefunction as
\begin{equation}
\vert \Psi\rangle =\prod_{i=1}^{N_{\rm cells}}\frac{1}{\sqrt{2}}\left(\alpha_i c^{\dagger}_{n_{i1},A}-\beta_i c^{\dagger}_{n_{i2},B}\right)
\end{equation}\noindent where $n_{i1}$ and $n_{i2}$ are two unit cells in the lattice, and $\alpha_i, \beta_i$ are complex numbers with unit modulus. The set of values $n_{i1}, n_{i2}, \alpha_i, \beta_i$ are random and depend on the details of the disorder configuration, but this general form has enough information to prove our result.
If we calculate the winding number of this ground state we find
\begin{equation}
\nu=\frac{1}{N_{\rm cells}}\sum_{i=1}^{N_{\rm cells}}\left(n_{i1}-n_{i2}\right).
\end{equation} Using the same ground state we can calculate the chiral displacement $2\langle \Gamma X\rangle$ and it is straightforward to find 
\begin{equation}
\frac{2\langle \Gamma X\rangle}{N_{\rm cells}}=\frac{1}{N_{\rm cells}}\sum_{i=1}^{N_{\rm cells}}\left(n_{i1}-n_{i2}\right)=\nu.
\end{equation}\noindent Hence, in the thermodynamic limit we expect that, even when the system is disordered, the chiral displacement will exactly match the winding number for this model. In the clean limit each dimer configuration contributes equally to the winding number and the chiral displacement, and as disorder is increased there are a distribution of configurations that always sum to an integer (assuming the system is not tuned to the critical point). 

\section{Dynamic extraction of the winding number from the mean chiral displacement}

To uncover how the winding number is encoded in the experimental system we will instead consider open boundary conditions. In this case a naive evaluation of the trace in Eq.~\eqref{realspacewindingnumber} yields identically zero, with the contribution from the bulk interior canceled exactly by the boundary modes. However, we can modify a formalism introduced by Bianco and Resta for a real-space calculation of the Chern number in quantum Hall insulators in Refs.~\cite{BiancoResta,BiancoResta2}, and which was subsequently used to describe Hofstadter quasicrystals in Ref.~\cite{TranDauphin}, to calculate the winding number instead. This method consists of defining a ``local topological marker'' that depends on the eigenfunctions of the system. This marker gives a local value for a topological invariant when evaluated in a region away from the physical boundary of the system. While this quantity is not exactly quantized, it converges smoothly and rapidly to the integer value of the corresponding invariant in the limit of an infinite system with mild assumptions of homogeneity of the bulk phase.

\begin{figure*}
	{\includegraphics[width=.95\columnwidth]{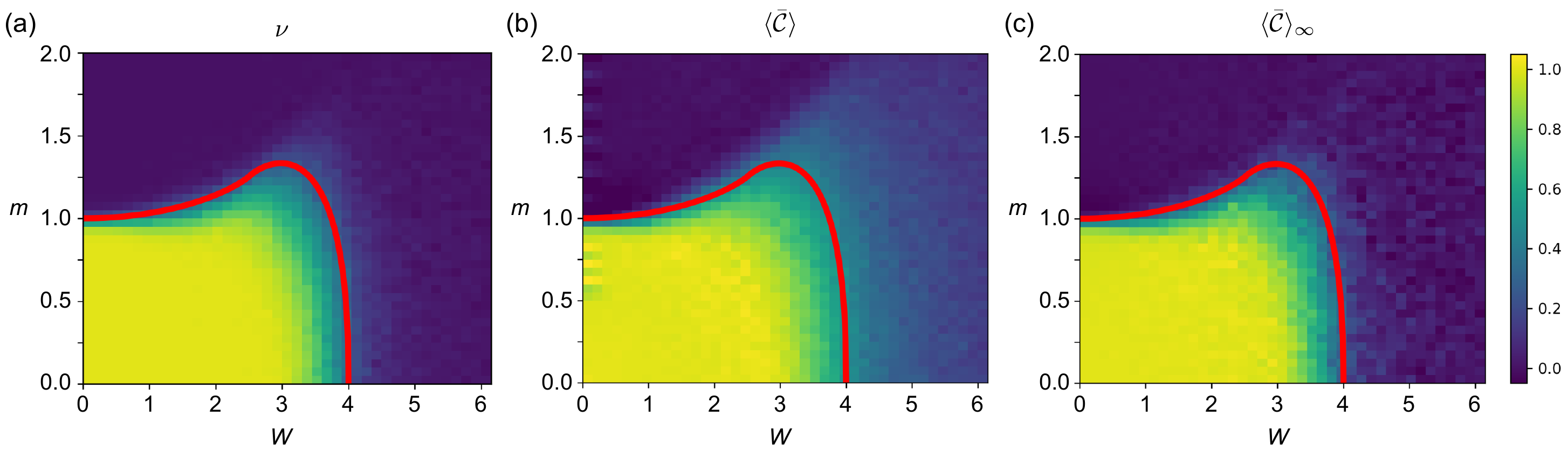}}
	\caption{\label{fig:winding_vs_mcd__phase_diagrams}
		Topological phase diagrams. \textbf{(a)} Real space winding number $\nu$,
		\textbf{(b)} the disorder- and time-averaged mean chiral displacement $\langle \bar{\mathcal{C}} \rangle$,
		and \textbf{(c)} the disorder-averaged MCD in the infinite time limit $\langle \bar{\C} \rangle_\infty$
		of the BDI model with disorder $W\equiv W_2=2W_1$. The simulations have been performed for a system of $50$ unit cells and for $1000$ disorder realizations. For the MCD, the sliding average has been done for times $\tau \in [5,50]$ with $\Delta \tau=1$. The red lines (identical in all panels) indicate the critical phase boundary, where the localization length diverges in the thermodynamic limit~\cite{DragonAIII}.
	}
\end{figure*}

In the following, we will use the idea of Bianco and Resta to compute the winding number $\nu$ in real space, which amounts to directly evaluating a symmetrized version of the argument of the trace per volume appearing in Eq.~\eqref{realspacewindingnumber} over the central part of the chain. Our topological marker then takes the form 
\begin{eqnarray}
\nu(j)&\equiv & \frac{1}{2}\left\{\left(Q_{BA}[X,Q_{AB}]\right)_{jA,jA}+\left(Q_{BA}[X,Q_{AB}]\right)_{jB,jB}+\left(Q_{AB}[Q_{BA},X]\right)_{jA,jA}+\left(Q_{AB}[Q_{BA},X]\right)_{jB,jB}\right\}\nonumber\\
&=&\frac{1}{2}\sum_{a=A,B}\left\{\left(Q_{BA}[X,Q_{AB}]\right)_{ja,ja}+\left(Q_{AB}[Q_{BA},X]\right)_{ja,ja}\right\}
\end{eqnarray}\noindent where $j$ indicates the lattice site index, and the subscripts $jA$ and $jB$ indicate the entries of the matrix corresponding to the $A$ or $B$ sublattice for lattice site $j.$ Note that this formula was symmetrized using the fact that the winding number can equivalently be written as $\nu=-\int_0^{2\pi}\frac{{\rm d}k}{2\pi}{\rm Tr}[(Q_{BA})^{-1}\,i\partial_k Q_{BA} ].$ To extract a value for the winding number in a disordered system we then average $\nu(j)$ over a small region ($\sim N/8$ unit cells) in the center of the lattice, and over disorder configurations. 
The numerical results of this method are shown for example in Figs.~\pref{fig2}{a} and \pref{fig:winding_vs_mcd__phase_diagrams}{a}. Their accuracy is confirmed by the fact that the numerical data display a very sharp drop in the vicinity of the red line, which corresponds to the set of tunneling ratios and disorder strengths where the localization length of the edge states diverges in an infinite system.

To make closer contact with the experimentally measured quantity, let us evaluate our topological marker at the 
central unit cell $j=0.$ Consider a basis of states  $\ket{0_a}=\sum_i \alpha_{ai}\ket{\phi_i}$
$(a=A,B)$ that are completely localized on such a cell, so that $X|0_a\rangle=0$. Here $\ket{\phi_i}$ are the energy eigenstates of the Hamiltonian $H$ satisfying $\ket{\phi_{-i}}=\Gamma\ket{\phi_i}$, and $E_{-i}=-E_i$ are the corresponding energies, ordered to satisfy $\ldots<E_{-2}<E_{-1}<0<E_1<E_2<\ldots$. To be explicit let \beq
M=\frac{Q_{BA}X Q_{AB} - Q_{BA}Q_{AB}X - Q_{AB}X Q_{BA} + Q_{AB}Q_{BA}X}{2},
\label{eq:symmetric}
\eeq\noindent be an operator such that $\nu(j)=\sum_{a=A, B}\langle j_a\vert M\vert j_a\rangle.$ For the center unit cell we can evaluate $\nu(0)$ using \begin{align}
\langle 0_a| M | 0_a \rangle
&= \frac{1}{2}\langle 0_a | Q_{BA}X Q_{AB} - Q_{AB}X Q_{BA} | 0_a \rangle
\end{align}\noindent where we have used $X\vert 0_a\rangle=0.$
We now use the identity $\Gamma_B Q = Q \Gamma_A $,\footnote{With open boundary conditions,  this equality does not hold for zero-energy edge states. But here we are only interested in a bulk state, like $|0\rangle$, which has negligible overlap with the edge states.} remember that $\Gamma_A$ and $\Gamma_B$ are projectors (so that, e.g., $\Gamma_A \Gamma_A=\Gamma_A$) and we can exploit the fact that the chiral operator is local (i.e., diagonal in the position basis), so that $[X,\Gamma_A]=[X,\Gamma_B]=0$, to find:
\beq
Q_{BA}X Q_{AB} - Q_{AB}X Q_{BA} 
= Q (\Gamma_A)^4 X Q - Q (\Gamma_B)^4 X Q =Q \Gamma X Q.
\eeq
Finally, using $Q=\mathbb{I}-2P_-$,
\begin{align}
\label{0QGXQ0}
\nu(0)=\sum_{a=A,B}\langle 0_a | M | 0_a \rangle
&= \sum_a\langle 0_a |\left[\frac{1}{2} \Gamma X - P_- \Gamma X - \Gamma X P_- + 2 P_-\Gamma X P_-\right]| 0_a \rangle  \nonumber\\ 
&= 2\sum_a\,\langle 0_a | P_-\Gamma X P_-| 0_a \rangle\nonumber\\  
&= 2\sum_a\left[ \sum_{i<0}|\alpha_{ai}|^2 \langle \phi_i | \Gamma X | \phi_i \rangle +
\sum_{i,j<0; i\neq j}\alpha_{ai}^*\alpha_{aj} \langle \phi_i | \Gamma X | \phi_j \rangle \right]
\nonumber\\  
&= \sum_a\left[\sum_{i}|\alpha_{ai}|^2 \langle \phi_i | \Gamma X | \phi_i \rangle
+  \sum_{i,j<0; i\neq j}\alpha_{ai}^*\alpha_{aj} \langle \phi_i | \Gamma X | \phi_j \rangle
+  \sum_{i,j>0; i\neq j}\alpha_{ai}^*\alpha_{aj} \langle \phi_i | \Gamma X | \phi_j \rangle\right].\nonumber\\
\end{align}
Numerically, we observe that the off-diagonal part of this expression provides a very small contribution (typically $\sim 1\%$ of the total), so that the sum is completely dominated by the diagonal term. 

Now we are ready to compare this expression to the chiral displacement calculated in experiment. The experimental 
procedure starts with an initial state localized on either $A$ or $B$ in the $j=0$ unit cell. The time evolution of this state is simply given by $e^{-iH\tau}\ket{0_a}$. As discussed in Ref.~\cite{Marianjp}, in a two-band model such as ours the mean chiral displacement at time $\tau$ can be defined as the mean value of the (time-evolved) operator $2\Gamma X$ over a single localized state: 
\begin{equation}
\label{eq:relwindmcd}
\mathcal{C}(\tau)=\bra{0_a} e^{iH\tau}(2\Gamma X) e^{-iH\tau}\ket{0_a}
=2\sum_{i} \vert \alpha_{ai} \vert^2 \bra{\phi_i} \Gamma X \ket{\phi_i}
+2\sum_{i\neq j}\alpha_{ai}^*\alpha_{aj} e^{-i (E_j-E_i)\tau}  \bra{\phi_i} \Gamma X \ket{\phi_j}.
\end{equation}
The second term of Eq.~\eqref{eq:relwindmcd} is rapidly oscillating, so that it converges to zero when averaged over sufficiently long time sequences.
The first term,
\beq
\label{Cinfty}
\langle \bar{\C} \rangle_\infty \equiv 2\sum_{i} \vert \alpha_{ai} \vert^2 \bra{\phi_i} \Gamma X \ket{\phi_i},
\eeq
is instead independent of time. 
Comparing with the dominant contribution to $\nu(0)$ in Eq.~\eqref{0QGXQ0} above, we see they differ in that $\langle \bar{\C} \rangle_\infty$ has an additional factor of 2 and is evaluated for a fixed $a=A$ or $B$, while $\nu(0)$ sums over both $A$ and $B.$ In the limit of zero disorder, it can be shown analytically that starting with the initial state with $a=A$ yields the exact same result for $\langle \bar{\C} \rangle_\infty$ as the case when $a=B.$ Hence, one can directly identify $\langle \bar{\C} \rangle_\infty=\nu(0)$ \cite{Marianjp}.
In the presence of disorder we find that, both numerically and experimentally, the  projections $\alpha_{ai}$ of the initial state on the different energy eigenstates are effectively randomly distributed variables so that (upon disorder average) the result from initializing on site $A$ is, on average, the same as initializing on site $B.$ Hence, one can still make the identification $\langle \bar{\C} \rangle_\infty=\nu(0)$ after disorder averaging.

\begin{figure*}
{\includegraphics[width=0.95\columnwidth]{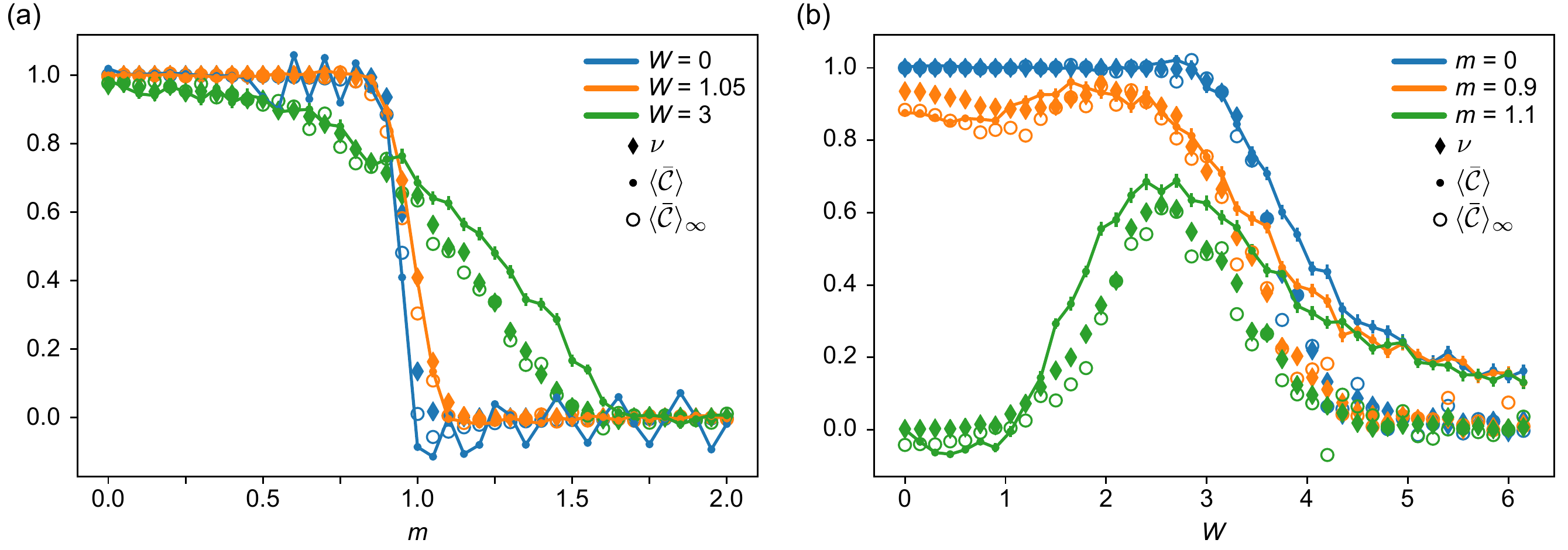}}
\caption{\label{fig:winding_vs_mcd}
Comparison of the winding number $\nu$ and the time- and disorder-averaged MCD $\langle \bar{\C} \rangle$.
\textbf{(a,b)} Cuts through the phase diagrams in Figs.~\pref{fig:winding_vs_mcd__phase_diagrams}{a} and \pref{fig:winding_vs_mcd__phase_diagrams}{b}, comparing the winding number $\nu$ (filled diamonds), the MCD $\langle\bar{\C}\rangle$ (lines with filled circles), and its infinite time limit $\langle\bar{\C}\rangle_\infty$ (open circles), for a BDI model with disorder ratio $W\equiv W_2=2W_1$.}
\end{figure*}

To conclude, we provide here a quantitative comparison of the various methods. Figure~\ref{fig:winding_vs_mcd__phase_diagrams} displays the phase diagram of the BDI model with disorder ratio $W\equiv W_2=2W_1$ obtained with the real space winding number [Fig.~\pref{fig:winding_vs_mcd__phase_diagrams}{a}] and with the MCD for both finite and infinite times [Figs.~\pref{fig:winding_vs_mcd__phase_diagrams}{b} and \pref{fig:winding_vs_mcd__phase_diagrams}{c}]. The simulations show results for a system with $50$ unit cells and have been averaged over 1000 disorder realizations. The sliding average for the MCD has been done between $\tau=5$ and $\tau=50$ with a sliding step $\Delta \tau =1$. The various subfigures show a very similar phase diagram, and all reproduce accurately the phase diagram expected for an infinite chain: the red line indeed represents the set of points where the localization length of the edge states is expected to diverge in the thermodynamic limit~\cite{DragonAIII}.

\begin{figure}
	\includegraphics[width=.5\columnwidth]{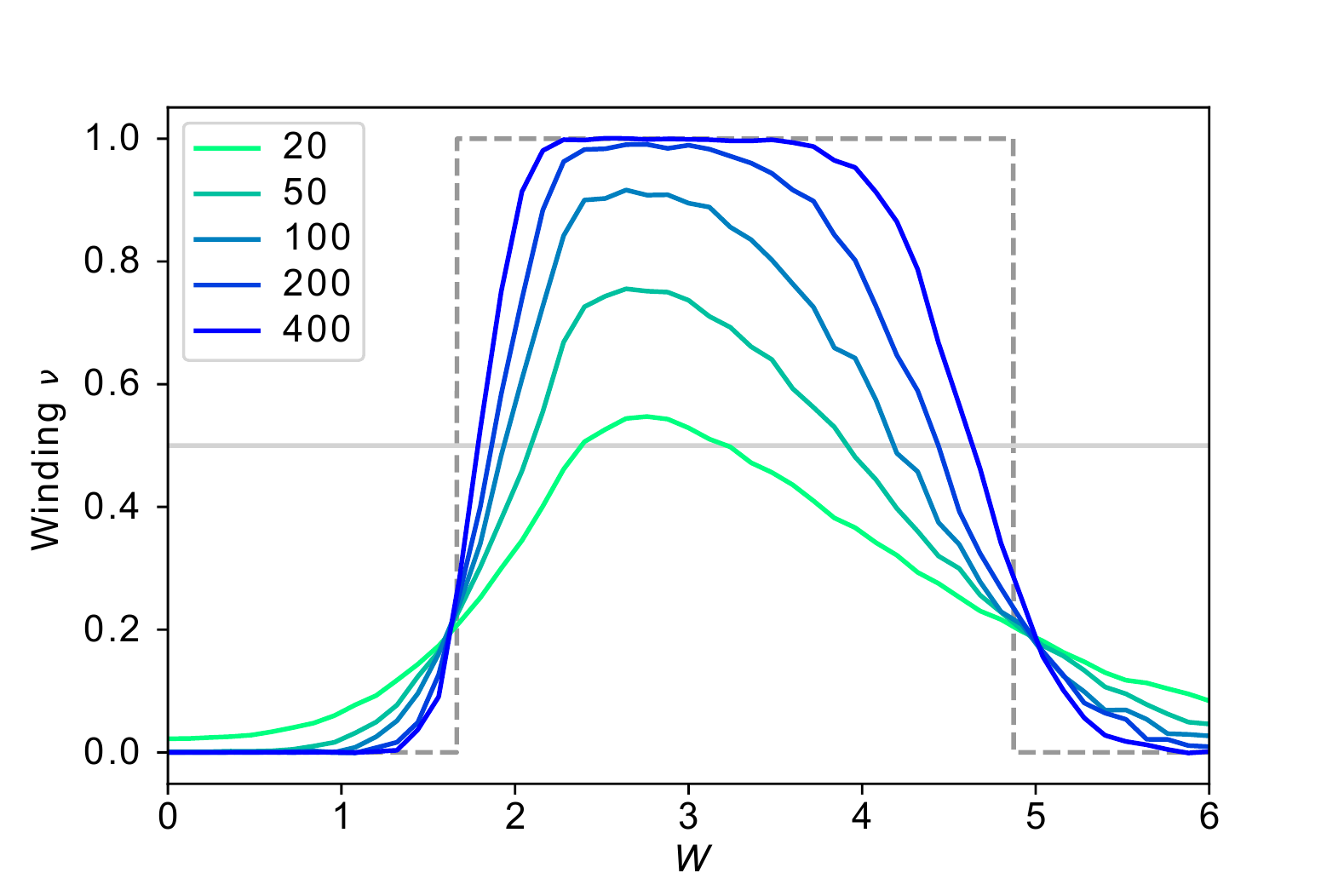}
	\caption{\label{fig:toplogicalAndersonPlateau}
		Emergence of the TAI plateau. 
		Winding number $\nu$ of the AIII model computed as a function of the disorder strength $W\equiv W_2$, with $W_1=0$ and $m=1.12$, averaged over 1000 disorder realizations. The various lines display results for systems with an increasing number of unit cells $N$, and the gray dashed line indicates the expected thermodynamic limit, given by the divergence of the localization length, as found in Ref.~\cite{DragonAIII}, and as indicated by the red line in Fig.~3(a) of the main text. 
	}
\end{figure}

For an even more accurate analysis, we compare the real space winding number and the MCD for several cuts in the phase diagram. Figure~\pref{fig:winding_vs_mcd}{a} shows the behavior of $\langle\bar{\C}\rangle$ (solid lines with filled circles) and $\langle\bar{\C}\rangle_\infty$ (open circles) for fixed values of the disorder $W$ and varying the hopping $m$. The results are in very good agreement with the real space winding number (filled diamonds). In this finite system size, the MCD has a sharp transition from $1$ to $0$ even for rather large disorder strength, like $W=1$. Figure~\pref{fig:winding_vs_mcd}{b} shows the behavior of the MCD for fixed values of $m$ and for varying $W$. When starting from the topological phase, the MCD presents a robust plateau before decreasing smoothly. The results coincide remarkably with the real space winding number and the sharpness of the transition here depends on the system size, as it is the case for the real space winding number. When starting from the trivial phase (e.g. $m=1.1$), the MCD starts from values close to $0$ and increases until reaching a small plateau before decreasing. This is the finite size signature of the topological Anderson insulator, since the sharp drops around a flat plateau, which one expects in the thermodynamic limit of infinite chains, are smoothed out on finite-size chains. 
The appearance of an actual large plateau at $\nu=1$ in the TAI region requires going to much larger system sizes, as shown in Fig.~\ref{fig:toplogicalAndersonPlateau}.

%